\documentclass[a4paper, 12pt]{report}
\usepackage[utf8]{inputenc}
\usepackage[T1]{fontenc}

\usepackage[dvips]{graphicx}

\usepackage{latexsym}
\usepackage{array}
\usepackage{amsmath}
\usepackage{amsfonts}
\usepackage{amssymb}
\usepackage{amsbsy}
\usepackage{amsthm}

\usepackage{dsfont}

\usepackage{float}
\usepackage{color}
\usepackage{times}

\newcommand{\mb}{\mathbf}
\newcommand{\cb}{\color{blue}}
\newcommand{\beq}{\begin{equation}}
\newcommand{\eeq}{\end{equation}}
\newcommand{\beqa}{\begin{eqnarray}}
\newcommand{\eeqa}{\end{eqnarray}}
\newcommand{\beqan}{\begin{eqnarray*}}
\newcommand{\eeqan}{\end{eqnarray*}}
\newcommand{\bit}{\begin{itemize}}
\newcommand{\eit}{\end{itemize}}
\newcommand{\nn}{\nonumber}
\newcommand{\beqna}{\begin{eqnarray}}
\newcommand{\eeqna}{\end{eqnarray}}
\newcommand{\bs}{\boldsymbol}
\newcommand{\nit}{\noindent}

\newcommand{\bfi}{\begin{figure}}
\newcommand{\efi}{\end{figure}}
\newcommand{\bc}{\begin{center}}
\newcommand{\ec}{\end{center}}

\usepackage[printonlyused]{acronym} 

\begin{document}
	
	\begin{titlepage} 
		\begin{center} 
			\vspace{0.4cm}
			{\Huge Technical Report:}\\
			\vspace{0.6 cm} 
			{\huge Time-Activity-Curve Integration in Lu-177 Therapies in Nuclear Medicine}\\
			\vspace{2.6 cm}
			\vspace{2.6 cm}

			{\Large from} \\\vspace{1.4 cm}  
			{\huge Dr. rer. nat. Theresa Ida Götz} \\\vspace{0.4 cm} 
			{\Large from Weiden (Germany)} \\
			\newpage
			\vspace{6.6 cm} 
		\end{center} 
	\end{titlepage}
	\newpage

\begin{acronym} 
	\setlength{\itemsep}{-\parsep} 
	\acro{spect/ct}[SPECT/CT]{{\em Single-Photon Emission Computed Tomography / CT} }
	\acro{pet/ct}[PET/CT]{{\em Positron-Emission Tomography / CT} }
	\acro{spect}[SPECT]{{\em Single-Photon Emission Computed Tomography} }
	\acro{pet}[PET]{{\em Positron-Emission Tomography} }
	\acro{ct}[CT]{{\em Computed Tomograpy} }
	\acro{mird}[MIRD]{{\em Medical Internal Radiation Dose Committee} }
	\acro{dvk}[DVK]{{\em Dose Voxel Kernel}  }
	\acro{dpk}[DPK]{{\em Dose Point Kernel}  }
	\acro{mc}[MC]{{\em Monte Carlo Simulation} }
	\acro{net}[NET]{{\em Neuroendocrine tumor} }
	\acro{adt}[ADT]{{\em Androgen Deprivation Therapy} }
	\acro{prrt}[PRRT]{{\em Peptide Radionuclide Receptor Therapy}  }
	\acro{mcspc}[mCSPC]{{\em metastativc Castration-Sensitive Prostate Cancer}  }
	\acro{sstr}[SSTR]{{\em Somatostatin Receptor}  }
	\acro{ttd}[TTD]{{\em Total Tumor Dose} }
	\acro{pc}[PC]{{\em Prostate Cancer}  }
	\acro{psma}[PSMA]{{\em Prostate Specific Membrane Antigen} }
	\acro{tac}[TAC]{{\em Time Activity Curve}  }
	\acro{em}[EM]{{\em Expectation Maximization} }
	\acro{tia}[TIA]{{\em Time Integrated Activity} }
	\acro{voi}[VOI]{{\em Volume Of Interest} }
	\acro{olinda}[OLINDA]{{\em Organ Level Internal Dose Assessment with Exponential Modeling} }
	\acro{se}[SE]{{\em Standard Error} }
	\acro{sd}[SD]{{\em Standard Deviation} }
	\acro{ss}[SS]{{\em Sample Size} }
	\acro{pf}[PF]{{\em Particle Filter} }
	\acro{snr}[SNR]{{\em Signal to Noise Ratio} }
	\acro{dnn}[DNN]{{\em Deep Neural Networks} }
	\acro{emd}[EMD]{{\em Empirical Mode Decomposition} }
	\acro{eemd}[EEMD]{{\em Ensemble Empirical Mode Decomposition} }
	\acro{meemd}[MEEMD]{{\em Multi-dimensional Ensemble Empirical Mode Decomposition}}
	\acro{bemd}[BEMD]{{\em Bi-Dimensional EMD} }
	\acro{cnn}[CNN]{{\em Convolutional Neural Network} }
	\acro{imf}[IMF]{{\em Intrinsic Mode Function} }
	\acro{bimf}[BIMF]{{\em Bi-dimensional Intrinsic Mode Function} }
	\acro{beemd}[BEEMD]{{\em Bi-dimensional \ac{eemd}} }
	\acro{git-bemd}[GiT-BEMD]{{\em ????} }
	\acro{bim}[BIM]{{\em Bi-dimensional Intrinsic Mode} }
	\acro{mle}[MLE]{{\em Maximum Likelihood Estimation} }
	\acro{relu}[ReLU]{{\em Leaky Rectified Linear Units} }
	\acro{mlp}[MLP]{{\em Multi-Layer Perceptrons} }
	\acro{lcn}[LCN]{{\em Local Contrast Normalization} }
	\acro{sgd}[SGD]{{\em Stochastic Gradient Descent} }
	\acro{kl}[KL]{{\em Kullback-Leibler divergence} }
	\acro{cga}[CgA]{{\em Chromogranin A level} }
	\acro{psa}[PSA]{{\em Prostate Specific Antigen} }
	\acro{auc}[AUC]{{\em Area Under the Curve} }
	\acro{sem}[SEM]{{\em State Evolution Model} }
	\acro{om}[OM]{{\em Observation Model} }
	\acro{pff}[PFF]{{\em Particle Fit} }
	\acro{sf}[SF]{{\em Simple Fit} }
	\acro{hu}[HU]{{\em Hounsfield Units} }
	\acro{sts}[STS]{{\em Soft Tissue Scaling} }
	\acro{cs}[CS]{{\em Center Scaling} }
	\acro{ds}[DS]{{\em Density Scaling} }
	\acro{ps}[PS]{{\em Percentage Scaling} }
	\acro{dvk-nn}[DVK-NN]{{\em DVK via Neural Networks} }
	\acro{nn}[NN]{{\em Neural Network} }
	\acro{de-nn-emd}[DE-NN-EMD]{{\em Dose Estimation via Neural Network and EMD} }
	\acro{de-nn}[DE-NN]{{\em Dose Estimation via Neural Network} }
	\acro{loocv}[LOOCV]{{\em Leave-One-Out Cross-Validation} }
	\acro{rlt}[RLT]{{\em Radioligand Therapy} }
	\acro{fbp}[FBP]{{\em Filtered Back-Projection}}
	\acro{sir}[SIR]{{\em Sampling Importance Resampling}}
	\acro{pca}[PCA]{{\em Principal Component Analysis}}
	\acro{cmsc}[CMSC]{{\em Comparable Minimal Scale Combination Principle}}
	\acro{gt}[GT]{{\em Ground truth}}
\end{acronym} 

\newpage
\chapter{Introduction}
Therapy with Lu-177-labelled somatostatin analogues and PSMA inhibitors are established tools in the management of patients with inoperable and metastasised neuroendocrine tumours (NET) and prostate cancer (PCa) \cite{Strosberg2017, Hofman2018}. Dosimetry could provide information for assessing the patient-individual dose-burden due to radiotherapy with Lu-177-DOTATOC and Lu-177-PSMA as radiopharmaceutical \cite{Zaknun2013}. Most commonly, the pharmaceutical is delivered over several (3-6) therapy cycles, where at the first cycle a dosimetry is performed. For this, several SPECT images can be collected at multiple time points. The resulting organ-wise time-activity-curve (TAC) is then integrated to estimate the cumulated activity, which, together with the so called S-value determines the energy dose, according to the Medical Internal Radiation Dose (MIRD) scheme. From this, the dose limiting organ is identified (often being kidneys), and the total amount of injected activity which can be applied over multiple subsequent radiotherapy cycles is determined.\\
Commonly, the dose assessment is limited to mean-organ doses, since per-voxel 3-D dose-distributions are often regarded as too noisy for estimating absorbed energy doses. Nevertheless, some studies from literature regarding this topic are available.

\section{Literature on Lu-177 Dosimetry}
In the following, a brief overview of the literature regarding dosimetry based on organ- or voxel-wise time-activity curves is provided:\\

Patient-specific dosimetry for I-131 tyroid tumor therapy is reported in \cite{Sgouros2004}. Such personalized dosimetry better accounts for radionuclide distribution and patient anatomy. However, its accuracy depends on the quality of the cumulated activity which enters the dosimetric calculations and which is derived from SPECT or PET imaging studies. Three to four image acquisitions were performed in $15$ patients with metastatic tyroid carcinomas. A voxel-wise integration over time of the PET images provided a 3D spatial distribution of accumulated activity, and absorbed dose maps were obtained employing the software package {\em 3D Internal Dosimetry}. The study indicated a high inter-tumor and inter-patient variability of absorbed dose distribution underlining the need for a patient-specific dosimetry.\\

A similar study is reported in \cite{Wilderman2006}, where the dosimetry in I-131 internal emitter therapy is discussed for follicular lymphoma patients being treated with I-131-tositumomab. The study employs voxel-wise integration of time-activity-curves derived from SPECT/CT images at multiple time points, which were subsequently registered at multiple time points on a dual-modality scanner. Maps of integrated TACs have been used together with CT images to determine absorbed energy dose spatial distributions by Monte Carlo (MC) simulations. Results are compared to state-of-the-art dosimetric calculations.\\  

A PRRT with Lu-177-labeled somatostatin analogues has been reported already by \cite{Wehrmann2007}. In their study, the authors compared the dosimetric parameter uptake, half-life (kinetics), and mean absorbed organ and tumor doses of Lu-177-DOTA-NOC and Lu-177-DOTA-TATE in $69$ patients suffering from a NET with high somatostatin receptor expression. Absorbed doses were assessed using the MIRD scheme. DOTA-TATE showed a higher uptake in tumor lesions as well as a higher mean absorbed dose. However, the high inter-patient variability deemed an individualized dosimetry obligatory.\\ 

Individualized dosimetry in patients undergoing therapy with Lu-177-DOTA-D-Phe-Ryr-octreotate was studied in \cite{Sandstrom2010}. SPECT data were analyzed considering whole organ volumes of interest (VOIs) to obtain the amount of radioactivity in the organ. In addition, small VOIs were employed to estimate the radioactivity concentration in the respective tissues. With a cohort of $24$ patients SPECT images were acquired at $1$h, $24$h, $96$h, and $168$h after infusion of the radio-pharmaceutical and absorbed doses were calculated in non-tumor affected organs like kidney, spleen and liver. Authors concluded that SPECT dosimetry based on small VOIs has been more reliable and delivered more accurate results than conventional planar dosimetry or dosimetry based on whole organ VOIs.\\

In a later study of Lu-177-labeled peptides for NET therapy, \cite{Kam2012} observed that side effects are few and remain mild, if kidney-protective agents were used, thus increasing the progression-free survival to more than $40$ months. Apart from kidneys, bone marrow is another organ at risk in radionuclide therapy.\\

In \cite{Sandstrom2013} an individualized dosimetry protocol for the bone marrow was developed. Furthermore,  patients,  undergoing fractionated therapy with $7.4$ GBq/cycle, were identified who first reached an accumulated dose of either $2$ Gy to the bone marrow or $23$ Gy to the kidneys. Planar whole-body and SPECT/CT images over the abdomen were acquired at $24$h, $96$h, and $168$h after administering Lu-177-octreotate. In addition, blood samples were collected six times during the first $24$h, with additional blood samples taken at $96$h and $168$h from several patients. Also urine samples were taken from several patients during the first $24h$. The absorbed radiation dose was computed from blood and urine activity curves as well as an organ-based analysis of whole body images. The absorbed dose to the kidney was calculated from the pharmacokinetic data obtained from SPECT/CT. The study revealed that the kidney was the dose limiting organ in $197$ out of $200$ patients. The number of cycles until the limiting dose was reached showed a high variability ($2 - 10$ cycles), thus suggesting the need for an individualized radiation treatment.\\

Another recent study by \cite{jackson2013automated} proposed an automatized voxel-wise dosimetry tool for radiotherapy based on multiple SPECT/CT images. The latter were acquired at $4, 24, 72$  hours  after Lu-177-DOTATATE PRRT for $17$ patients suffering from advanced NETs. Deformable image registration and in-house developed algorithms for pharmacokinetic uptake interpolation at voxel level were employed. Absorbed energy dose was computed from cumulated activity based on MC-determined voxel S-values and compared to results obtained using a conventional MIRD protocol (OLINDA).\\

Dosimetry of organs and tumors is often avoided because of the additional burden for patients. Therefore, in  \cite{hanscheid2018dose} an approximation has been advocated which allows the calculation of absorbed energy doses from a single measurement of the abdominal radioactivity distribution. In the study, the activity kinetics were assessed in $29$ patients suffering from NETs or meningioma after administering Lu-177-DOTATATE or -DOTATOC. Mono-or bi-exponential decay curves were fit to SPECT/CT imaging data from kidneys, livers, spleens and NET lesions. The time-integrated activity was compared to an approximation given by the product of three factors: the time $t$ of a single measurement, the expected reading at time $t$ and the factor $2/ \ln 2$. Tissue-specific deviations of the approximation from the time-integrated activity were computed for several time points ranging from $t = 24$h to $t = 144$h. Results were inconclusive in that best accordance was achieved for $t = 96$h, while a reduced accuracy resulted for shorter and longer time spans.

\color{black}
\chapter{Dataset}

\section{Patient collectiv}

\begin{table}[h!]
	\label{tab:pat_net}
	\caption{Patient data concerning the DOTATOC radiotherapy. Here $m$ denotes the weight of each person, $A_0$ the initially injected radioactivity, $g$ the grading of the \ac{net}, $CgA_1$ denotes the \ac{cga} before therapy and $CgA_2$ after therapy, $\Delta t$ the time betweent the $CgA$ values, $n$ the number of therapy cycles and the sex is coded in $1$ for female and $0$ for male.}
	\vspace{3mm}
	\begin{center}
		\begin{tabular}{c|c|c|c|c|c|c|c|c}
			Age 	& sex		& $g$	& $m$	    & $A_0$ &	 $CgA_1$  & $CgA_2$ 	&	$\Delta t $  	& $n$\\
			$[y]$	& 		& 	&  $(kg)$ 	    & $(MBq)$ &	 ($\mu g/l$)  & ($\mu g/l$)	&	$(m)$  	& \\ \hline
			45			& 1	& 3	& 64			& 6077		&   	$494$			&	$1026$			&	$5$						&	$4$				\\
			66			& 0		& 3	& 79			& 6375		&   	$1121$			&	$706$			&	$4$						&	$4$				\\
			54			& 0		& 1	& 77			& 7054		&   	$661$			&	$1956$			&	$4$						&	$2$				\\
			67			& 1	& 2	& 70			& 6821		&   	$1352$			&	$1473$			&	$6$						&	$3$				\\
			78			& 0		& 2	& 88			& 5773		&   	$34$			&	$31$			&	$4$						&	$3$				\\
			54			& 0		& 2	& 56			& 7023		&   	$54$			&	$77$			&	$4$						&	$3$				\\
			67			& 0		& 2	& 75			& 5809		&   	$597$			&	$1748$			&	$2$						&	$2$				\\
			54			& 0		& 2	& 71			& 6644		&   	$1319$			&	$682$			&	$4$						&	$4$				\\
			77			& 1	& 2	& 49			& 6378		&   	$224$			&	$246$			&	$1$						&	$1$				\\
			52			& 0		& 3	& 78			& 6233		&   	$46$			&	$83$			&	$5$						&	$2$				\\
			52			& 0		& 3	& 103			& 7265		&   	$76$			&	$47$			&	$4$						&	$3$				\\
			71			& 0		& 2	& 118			& 6757		&   	$51$			&	$51$			&	$3$						&	$2$				\\
			49			& 0		& 3	& 80			& 6328		&   	$139$			&	$214$			&	$2$						&	$2$				\\ \hline\hline
			$66.5$		& 			& $2.3$	& $77.5$	& 6503		&		$589.8$			&	$768.1$			& 	$3.9$					&	$2.8$				\\
		\end{tabular}
	\end{center}
\end{table}

This study comprises $26$ patients suffering from \ac{net}s ($13$) or \ac{pc} ($13$) and which underwent a $^{177}Lu$-DOTATOC or $^{177}Lu$-\ac{psma} therapy. The patient cohort consists of $23$ male and $3$ female patients with an average age of $63.8\pm 10$ years at time of therapy. The on average injected activity was $A_0 = 6643 \pm 421$ MBq.

The information about the patients suffering on \ac{net} is collected in table \ref{tab:pat_net} and for the patients with prostate cancer in \ref{tab:patient_data_psa}.

\begin{table}
\centering
\caption{Patient data concerning the \ac{psma} radiotherapy. Here $m$ denotes the weight of each person, $A_0$ the initially injected radioactivity, $Gleason$ the Gleason Score of the \ac{pc}, $PSA_1$ denotes the \ac{psa} level before therapy and $PSA_2$ after therapy, $\Delta t$ the time betweent the $PSA$ values and $n$ the number of therapy cycles.}
\label{tab:patient_data_psa}
\vspace{3mm}
\begin{tabular}{c|c|c|c|c|c|c|c}
	Age 	& Gleason 		& $m$	    	& $A_0$ 	&	 $PSA_1$   & $PSA_2$ 	&	$\Delta t$  & $n$\\
	 $[y]$	& 		  		& $(kg)$		& $(MBq)$ 	&	 ($ng/ml$) & ($ng/ml$)	&	$(m)$ 	 	& \\ \hline
   80		& 	$9$			& 88			& 6699		&   	$1277$			&	$594$			&	$2$						&	$1$		\\
   67		& 	$8$			& 104			& 6997		&   	$96$			&	$120$			&	$6$						&	$2$	\\
   75		& 	$9$			& 74			& 6314		&   	$22$			&	$5$				&	$2$						&	$1$	\\
   55		& 	$6$			& 113			& 5931		&   	$122$			&	$523$			&	$2$						&	$1$	\\
   61		& 	$9$			& 85			& 6899		&   	$32$			&	$74$			&	$5$						&	$2$	\\
   66		& 	$9$			& 97			& 7139		&   	$92$			&	$199$			&	$2$						&	$1$	\\
  63		& 	$9$			& 94			& 7103		&   	$484$			&	$289$			&	$2$						&	$1$	\\
   65		& 	$9$			& 80			& 7030		&   	$58$			&	$0.1$			&	$3$						&	$2$	\\
   66		& 	$9$			& 86			& 6545		&       $469$			&	$445$			&	$1$						&	$1$	\\
   78		& 	$9$			& 106			& 6768		&   	$93$			&	$86$			&	$3$						&	$2$	\\
   79		& 	$9$			& 75			& 6915		&   	$138$			&	$25$			&	$2$						&	$1$	\\
   62		& 	$8$			& 68			& 6824		&   	$2251$			&	$1946$			&	$2$						&	$1$	\\
   55		& 	$9$			& 94			& 7006		&   	$148$			&	$53$			&	$2$						&	$2$	\\ \hline \hline
	$65.7$	&	$8.4$		& $90$			& $6782$&		$406.3$			&	$336.0$			& 	$2.6$					&	$1.4$
\end{tabular}
\end{table}

\section{Image Acquisition}

\subsection{\ac{spect}}  
Data was acquired on a \ac{spect/ct} system (Siemens Symbia T2) at time points $4$h, $24$h, $48$h, and $72$h after administering the radiopharmaceutical. The acquisition of the $24$ h p.i. image was carried out as part of a hybrid \ac{spect/ct} acquisition, After manual co-registration, the \ac{ct} part was used for attenuation correction of the subsequent \ac{spect}. Acquisition was done based on an inhouse standard quantitative $^{177}Lu$ - protocol, which is described in detail in \cite{Sanders2014}. Therefore, it is herein only briefly outlined:

\begin{itemize}
\item \ac{spect} using medium energy collimators, $3^{\circ}$ angular sampling, $15$ min total dwell time
\item Iterative ordered-subset conjugate-gradient reconstruction of the $208$ keV photopeak data with $24$ iterations, $1$ subset, matrix $256$x$256$
\item Point-spread-function modelling in reconstruction
\item Triple energy window based scatter correction
\item \ac{ct}-based attenuation correction
\item No post-reconstruction smoothing
\end{itemize}

The reconstruction, using the ordered-subset conjugate-gradient algorithm, was carried out on a Siemens research workstation. The algorithm outputs fully quantitative reconstructed \ac{spect} images providing activities per voxel measured in $[Bq/ml]$. Furthermore the \ac{spect} data is alined with the corresponding \ac{ct} data. 

\subsection{\ac{ct}}
Additionally, a low-dose \ac{ct} was carried out as part of the multi-modal \ac{spect/ct} acquisition to enable necessary attenuation corrections of the \ac{spect} images. The \ac{ct} covered the same field-of-view as the \ac{spect} and was acquired and reconstructed using the following parameters:

\begin{itemize}
\item Slice collimation of 2 x 5 mm, pitch of 1.8, time per rotation of 0.8 s, tube voltage of 130 kVp, tube current of 30 mAs effective
\item Filter-Back-Projection reconstruction with B08s and B41s Kernels, $512$x$512$ matrix, $2.5$ mm slice thickness
\item B08s image was used for attenuation correction of the \ac{spect} data
\item B41s image was used for defining organ and tumor volumes-of-interest
\end{itemize}

Reconstructed \ac{spect} images were partitioned into voxels of size $V_{vox} = 4.79\ mm^3$ and were cropped to $82^3$ voxels. Thereby large portions of the reconstructed volume were removed which predominantly contained air.
\section{Common Preprocessing} 
\label{chap:Common Preprocession}

\begin{figure}[!h]
	\centering
	\includegraphics[width=0.9\textwidth]{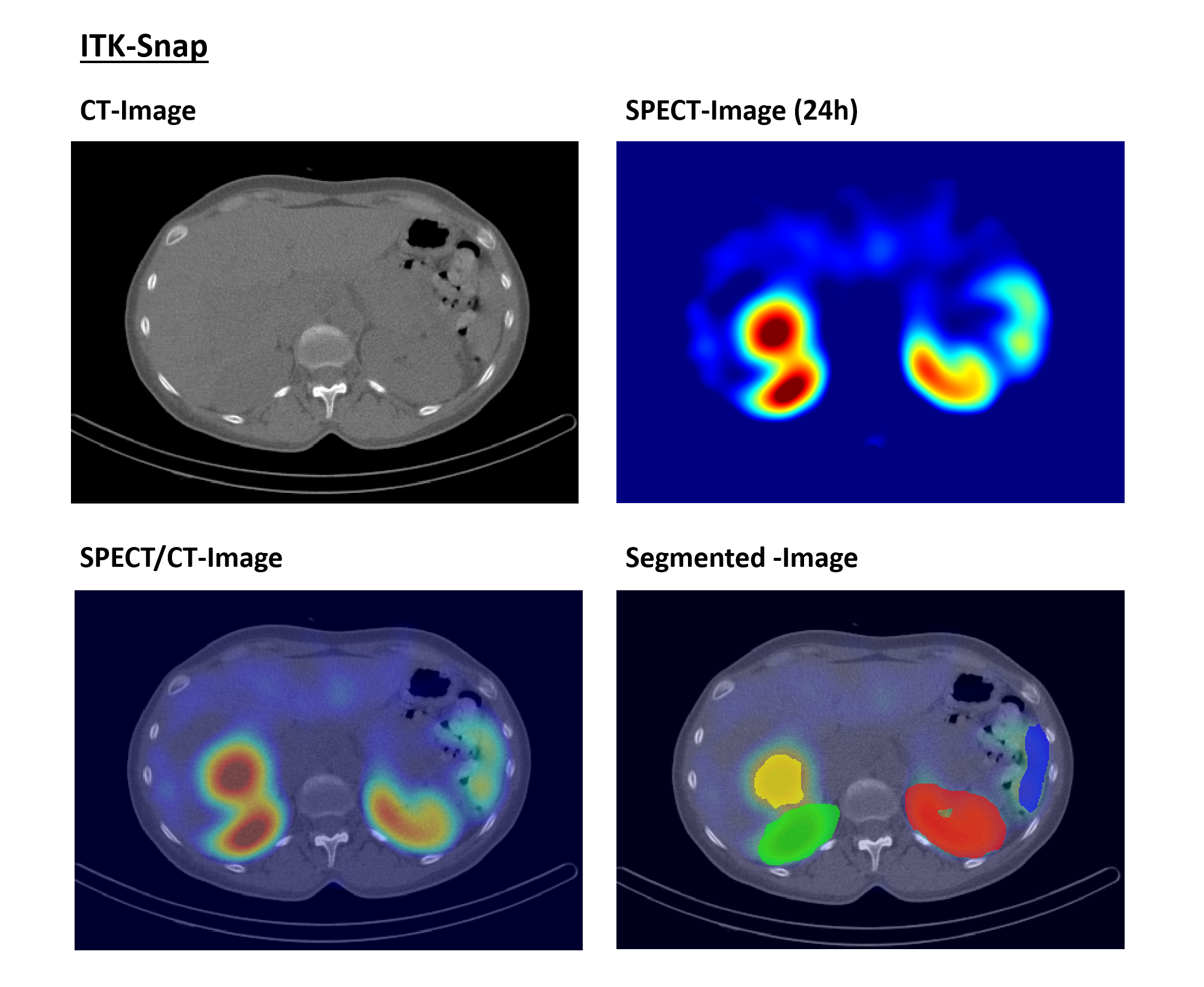}
	\caption{\ac{ct}-image, \ac{spect}-image and hyprid \ac{spect/ct}-image as shown in ITK-snap. In the right bottom corner the segmentation of the right and left kidney, the spleen and a liver tumor can be seen.}
	\label{fig:segments}
\end{figure}

In figure \ref{fig:segments} a \ac{ct}-image, a \ac{spect}-image and a \ac{spect/ct}-image are presented. For all patients, the kidneys, spleen and tumors were segmented by a medical expert. The tumors were sorted to bone lesion, lymph nodes, visceral and primary tumor. The segmentation was done using the software ITK-snap \cite{py06nimg}, which was carried out on the original \ac{ct} datasets. The resulting segmentation was subsequently down-sampled to match the voxel-size of $(4.7mm)^3$ of the \ac{spect} image. For each patient, the follwoing information is available:

\bit
\item \ac{spect}-Image acquired 4h after injection
\item \ac{spect}-Image acquired 24h after injection
\item \ac{spect}-Image acquired 48h after injection
\item \ac{spect}-Image acquired 72h after injection
\item \ac{ct}-Image acquired 24h after injection 
\item Organ and tumor segmentation
\eit 

\begin{figure}[!htb]
	\centering
	\includegraphics[width=0.9\textwidth]{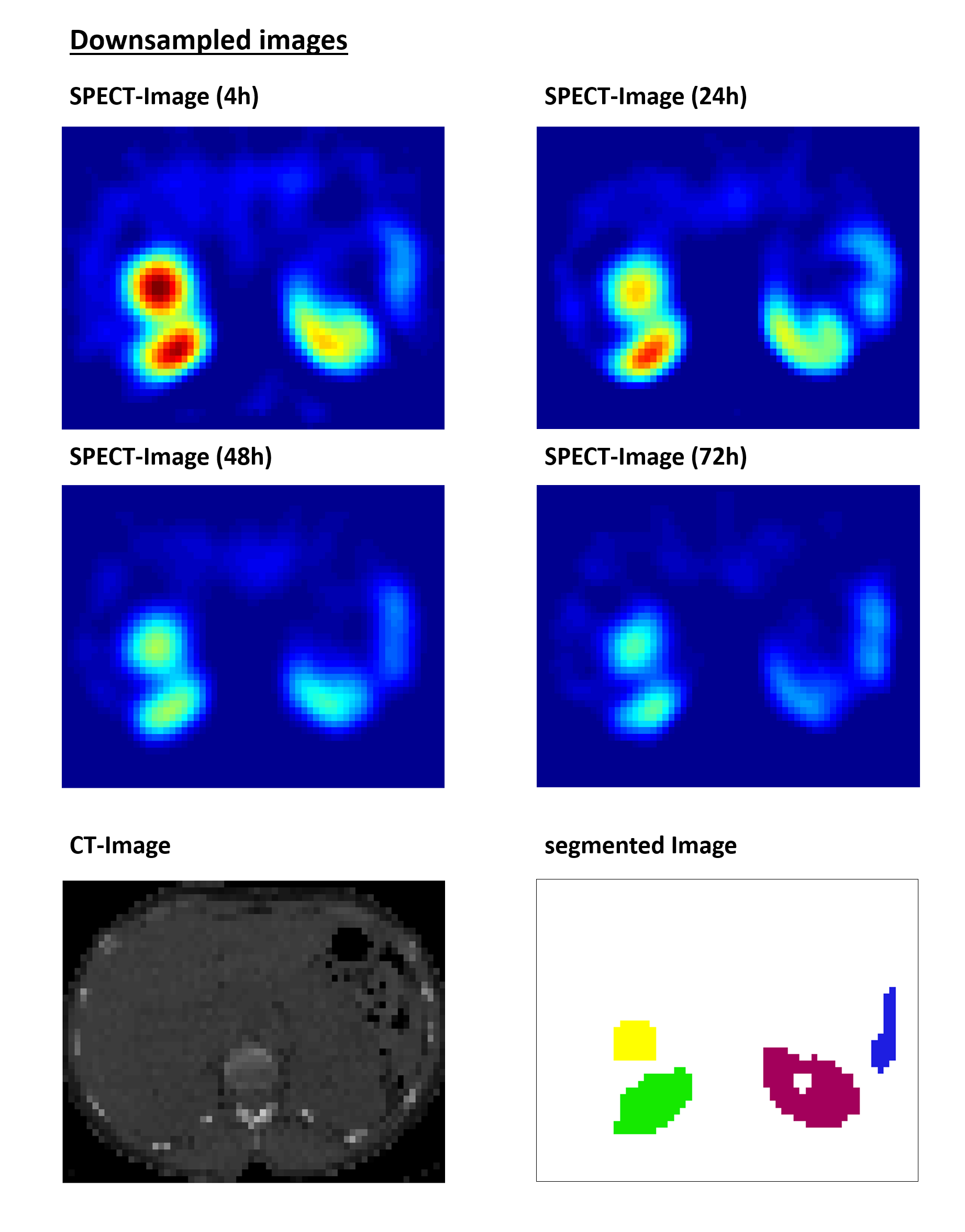}
	\caption{The four \ac{spect}-images, the \ac{ct}-image and the corresponding segmentation map illustrated for one slice of one patient.}
	\label{fig:patient_dat}
\end{figure}

Exemplarly, for one patient and one slice the eight images are illustrated in figure \ref{fig:patient_dat}.
		
\chapter{Sequential Monte Carlo Techniques}
\label{PF}

Non-linear dynamical systems evolve either continuously or via transitions between discrete states. The latter approximation is only reasonable if the mean residence time in any state is longer than the transition time. Particle filters represent discrete state space models which estimate the states $\mb{X}(\mb{r},t)$ of a dynamical system from noisy and disturbed observations  $\mb{Z}(\mb{r},t)$. This is achieved by combining available measured data $\mb{Z}(\mb{r},t)$ with prior knowledge about the underlying physical phenomena driving the dyanmics and the measurement process. The underlying physics allow to construct a state evolution model while the known measurement process provides information to construct an observation model. With these ingredients and observations of the system dynamics,  estimates of the unobservable dynamic state variables can be produced sequentially.

\section{Background}
State evolution models are commonly based on a Markov model, while observation models need to reflect the specifics of the physical measurement process. Both, evolution and observation models are most often described in terms of conditional probability density functions. Hence we have (throughout we use the short hand notation $\mb{z}(t_k) \equiv \mb{z}_k, \{\mb{z}(t_0), \ldots , \mb{z}(t_k)\} \equiv \mb{z}_{0:k}$ etc.)

\beqna
\mb{x}_{k+1} &\sim& p\left( \mb{x}_{k+1} | \mb{x}_{0:k} \right) \approx p\left( \mb{x}_{k+1} | \mb{x}_{k} \right) \\
\mb{z}_k         &\sim& p\left(\mb{z}_k | \mb{x}_{0:k} \right) \approx p\left(\mb{z}_k | \mb{x}_{k} \right)
\eeqna 

\noindent where $\sim$ means ``represented through'', $\mb{x}_k$ denotes an unobservable state vector of the dynamical system and $\mb{z}_k$ a corresponding observation vector of the system through a measurement at discrete time $t_k$  \cite{MacKay2003}. Note that the conditional probability densities only depend on the last state of the system. This is in accord with the Markov assumption which expresses common intuition that any prediction of a future state should be based primarily on the recent history of the evolving system. 

Now given the task that we want to predict unobservable states of an evolving system from sequential noisy observations, Bayesian inference \cite{Kendall1999} allows to integrate available information, i.~e. observations $\mb{z}_{0:k}$ and an associated physical model of the dynamical system, parametrized through dynamic states $\mb{x}_{0:k}$.  As new information $\mb{z}_k$ is available, it can be combined with previous information $p(\mb{x}_{0:k-1},\mb{z}_{0:k-1})$ via the Bayes Theorem to deduce the underlying system state $\mb{x}_k$. Employing the Markov assumption we have

\beq 
p(\mb{x}_k | \mb{z}_{0:k}) =  p(\mb{x}_k | \mb{z}_k,\mb{z}_{0:k-1}) 
\approx  \frac{p(\mb{z}_{k} | \mb{x}_{k})p(\mb{x}_k | \mb{z}_{k-1})}{\sum_k \left( p(\mb{z}_k | \mb{x}_k) p(\mb{x}_k | \mb{z}_{k-1}) \right)} 
\eeq  

The latter forms the basis for statistical estimates of the  posterior density $p(\mb{x}_k | \mb{z}_{0:k})$ of the unobservable state vector $\mb{x}_k$ given the likelihood of the observation $p(\mb{z}_{k} | \mb{x}_{k})$, some prior knowledge $p(\mb{x}_k | \mb{z}_{k-1})$ about the observation process and the data evidence $p(\mb{z}_k)$. 

Note that the term posterior density denotes different densities depending on the application: 

\bit
\item
Filtering: $p(\mb{x}_k | \mb{z}_{0:k})$
\item
Prediction: $p(\mb{x}_{k+\tau} | \mb{z}_{0:k})$
\item
Smoothing: $p(\mb{x}_{k-\tau} | \mb{z}_{0:k})$
\eit

\noindent where $\tau > 0$ denotes the prediction or smoothing lag.

The most widely known Bayesian filter method is the  Kalman filter \cite{Kalman1960}, which, however, relies on linear models with additive Gaussian noise, i.~e. it assumes the distribution $p(\mb{x}_k | \mb{z}_{0:k})$ to be a multi-variate Gaussian. Extensions of the Kalman filter concern sequential Monte Carlo methods \cite{Fishman1996} which represent the posterior density in terms of random samples (particles) and associated weights. Such sequential Monte Carlo methods are variously known as {\em Particle Filter}, bootstrap filter \cite{Gordon1993}, survival of the fittest \cite{Kanazawa1995}, condensation algorithm \cite{Isard1998} and can be applied to non-linear models with non-Gaussian errors \cite{Doucet2001}. Thus, given any dynamical state representation $f(\mb{x}_k)$, 
one can estimate the related expected value via

\beqa 
\mathbb{E}(f(\mb{x}_k)) &=& \int f(\mb{x}_k)p(\mb{x}_k | \mb{z}_{0:k}) d\mb{x}_k \nn \\
&=& \frac{\int f(\mb{x}_k) p(\mb{z}_k | \mb{x}_k)p(\mb{x}_k | \mb{z}_{k-1}) d\mb{x}_k }{\int p(\mb{z}_k | \mb{x}_k )p(\mb{x}_k | \mb{z}_{k-1}) d\mb{x}_k} \nn \\
&\approx & \sum_{l=1}^L w_k^{(l)}f(\mb{x}_k^{(l)})
\eeqa 

\nit Here $\mb{x}_k^{(l)}$ denotes a set of samples, called {\em particles}, drawn from the posterior distribution $p(\mb{x}_k | \mb{z}_{k-1})$. The sampling weights are given by

\beq 
w_k^{(l)} = \frac{p \left( \mb{z}_k | \mb{x}_k^{(l)} \right)}{\sum_{l'=1}^L p \left( \mb{z}_k | \mb{x}_k^{(l')} \right)}
\eeq 

\nit where the same samples are used in the numerator and the denominator. Hence a particle filter approximates the posterior distribution $p(\mb{x}_k | \mb{z}_{0:k})$ by a set of particles  $\mb{x}_k^{(l)}$, sampled from $p(\mb{x}_k | \mb{z}_{k-1})$, and their weights $\{ w_k^{(l)}\}$. The latter are normalized to $0 \le w_k^{(l)} \le 1$ and obey the closure relation $\sum_l w_k^{(l)} = 1$.

In practice one is concerned with a sequential sampling scheme, which works as follows:

\bit
\item 
Suppose, a set of samples $\{ \mb{x}_{k}^{(l)} \}$ and related weights $w_k^{(l)}$ have been obtained at time $t_k$. These samples and weights represent the posterior distribution $p(\mb{x}_k | \mb{z}_{k})$ at time step $t_k$.
\item 
Suppose, an observation $\mb{z}_{k+1}$ has been made subsequently
\eit

How can one estimate corresponding samples $\mb{x}_{k+1}^{(l)}$ and weights $w_{k+1}^{(l)}$?

\bit 
\item 
First, new samples are drawn from the distribution 
\beqa  
p(\mb{x}_{k+1} | \mb{z}_{0:k}) &=& \int p(\mb{x}_{k+1} | \mb{x}_k, \mb{z}_{0:k}) p(\mb{x}_k | \mb{z}_{0:k}) d\mb{x}_k \nn \\
&=& \int p(\mb{x}_{k+1} | \mb{x}_k) p(\mb{x}_k | \mb{z}_{0:k}) d\mb{x}_k \nn \\
&=& \int p(\mb{x}_{k+1} | \mb{x}_k) p(\mb{x}_k | \mb{z}_{k}, \mb{z}_{0:k-1}) d\mb{x}_k \nn \\
&=& \frac{\int  p(\mb{x}_{k+1} | \mb{x}_k) p(\mb{z}_k | \mb{x}_{k}) p(\mb{x}_k | \mb{z}_{k-1}) d\mb{x}_k }{\int p(\mb{z}_k | \mb{x}_k ) p(\mb{x}_k | \mb{z}_{k-1}) d\mb{x}_k} \nn \\
&=& \sum_{l=1}^L w_k^{(l)} p\left( \mb{x}_{k+1} | \mb{x}_k^{(l)} \right)
\eeqa 

where the conditional independence property has been used

\beqa 
p(\mb{x}_{k+1} | \mb{x}_k, \mb{z}_{0:k}) &=& p(\mb{x}_{k+1} | \mb{x}_k) \\
p(\mb{z}_{k} | \mb{x}_k, \mb{z}_{0:k-1})  &=& p(\mb{z}_{k} | \mb{x}_k)
\eeqa 

The above distribution is a mixture distribution. Hence, samples $\mb{x}_k^{(l)}$ can be drawn by choosing a mixture component $l$ with probability $w_k^{(l)}$ and drawing a sample from the corresponding mixture component of the distribution.

\item 
Second, for each sample the new observation $\mb{z}_{k+1}$ is used to estimate the related weights

\beq 
w_k^{(l)} \propto p(\mb{z}_{k+1} | \mb{x}_{k+1}^{(l)})
\eeq 
\eit 

Thus using a probabilistic description and the Markov assumption, one is concerned with the following densities:

\bit
\item
The  likelihood of the state $p(\mb{x}_k | \mb{z}_{k-1})$ 
\item
The Markov transition model $p( \mb{x}_{k+1} | \mb{x}_k)$ $\rightarrow$ state evolution model
\item
The sensor model $p(\mb{z}_{k} | \mb{x}_k) $ $\rightarrow$ observation model
\eit 
\section{State Estimation}

While theoretical derivations are based on the general Markov model, algorithms are based on evolution and observation models as detailed next.

\subsection{The model equations}
Let the dynamic state of a system be described by a state vector $\mb{x}(t) \in \mathcal{R}^N$ whose components represent the dynamical variables of the system. The dynamic evolution of the system, called the state evolution model $\mb{f}(...)$, is described by the following equation for the state vector 

\beq  
\mb{x}(t_k) \equiv \mb{x}_k = \mb{f} \left( \mb{x}_{k-1}, \bs{\epsilon}^{(x)} \right) 
\eeq 

\noindent where $k = 1, 2, ....... $ denotes discrete time instances $t_k$ and $\bs{\epsilon}^{(x)} \in \mathcal{R}^L$ denotes the state noise.

If measurements are available at $t_k$, they are collected in the observation vector $\mb{z}$, and its relation to the state vector is given by the observation model $\mb{h}(...)$

\beq  
\mb{z}(t_k) \equiv \mb{z}_k = \mb{h} \left( \mb{x}_k, \bs{\epsilon}^{(m)} \right)
\eeq 

\noindent where $\bs{\epsilon}^{(m)} \in \mathcal{R}^M$ denotes the measurement noise. The various noise distributions $p(\bs{\epsilon}^{(x)})$ and  $p(\bs{\epsilon}^{(m)})$ are assumed to be known. In some applications, also additional known {\em control variables} $\mb{u}_k$ enter the problem, but are omitted here.

The state estimation problem aims at obtaining information about the state vector $\mb{x}_k$ based on the state evolution model $\mb{f}(...)$ and based on the measurements $\mb{z}_{1:k}$, given by the observation model $\mb{h}(...)$.

\subsection{The simplifying assumptions}
The state evolution and observation models rely on the following simplifying assumptions:

\bit
\item
The sequence of state vectors $\mb{x}_k$ forms a Markov process \cite{Eberle2017} and does not depend on the sequence of observations

\beq  
p(\mb{x}_k | \mb{x}_{0:k-1}, \mb{z}_{1:k-1})  = p(\mb{x}_k | \mb{x}_{k-1})
\eeq 

\item 
The observations $\mb{z}_k$ form a {\em Markov process} with respect to the history of the state vector $\mb{x}_k$, i.~e.

\beq  
p(\mb{z}_k | \mb{x}_0, \mb{x}_1, \mb{x}_2, \ldots , \mb{x}_{k-1}) = p(\mb{z}_k | \mb{x}_{k})
\eeq 

\item
The probability of an observation only depends on the current state

\beq 
p(\mb{z}_k | \mb{x}_{0:k}, \mb{z}_{1:k-1}) = p(\mb{z}_k | \mb{x}_k)
\eeq 

\item
For $i \neq j$ the noise vectors $\bs{\epsilon}^{(x)}_i$ and $\bs{\epsilon}^{(x)}_j$, as well as $\bs{\epsilon}^{(m)}_i$ and $\bs{\epsilon}^{(m)}_j$, are mutually independent

\item
The noise vectors $\bs{\epsilon}^{(x)}_i$ and $\bs{\epsilon}^{(m)}_j$ are mutually independent for all $i,j = 1, 2, ....$

\item
All noise vectors are also mutually independent of the initial state $\mb{x}_0$ 
\eit


\subsection{The problems to be solved}

Several problems can be tackled with this approach

\bit
\item 
The {\em prediction problem} determines $p(\mb{x}_k | \mb{z}_{1:k-1})$, i.~e. it predicts the state of the system at time $t_k$ based on observations made at times $t_1, \ldots , t_{k-1}$

\item
The {\em filtering problem} determines $p(\mb{x}_k | \mb{z}_{1:k})$, i.~e. it filters out from the observations made up to time point $t_k$ the state of the system at time $t_k$.

\item
The {\em fixed lag smoothing problem} determines $p(\mb{x}_k | \mb{z}_{1:k+\tau})$, i.~e. it determines the state of the system at time $t_k$ given observations over a time span $t_1, \ldots , t_{k+\tau}$, where $\tau \ge 1$ denotes the fixed time lag.

\item 
The {\em whole domain smoothing problem} determines $p(\mb{x}_k | \mb{z}_{1:K})$, where $\mb{z}_{1:K} = \mb{z}_i, i = 1, \ldots , K$ denotes the complete observation sequence.
\eit

In this study, only the {\em filtering problem} is of interest here. By assuming that $p( \mb{x}_0 | \mb{z}_0) = p( \mb{x}_0)$ is available, the {\em posterior probability density} $p( \mb{x}_k | \mb{z}_{1:k})$ is then obtained with Bayesian filters in two steps: {\em prediction} and {\em update}.

\section{Bayesian Filters}

The Bayesian solution to compute the {\em posterior distribution}, $p(\mb{x}_k | \mb{z}_{1:k})$, of the state vector $\mb{x}(t_k)$, given past observations $\mb{z}_{1:k}$, is given by the general {\em Bayesian update recursion}

\beqna 
\mbox{measurement update}: p\left( \mb{x}_k | \mb{z}_{1:k}\right) &=& \frac{p(\mb{z}_k | \mb{x}_k)p(\mb{x}_k | \mb{z}_{1:k-1})}{p(\mb{z}_k | \mb{z}_{1:k-1})} \\
\mbox{normalization}: p(\mb{z}_k | \mb{z}_{1:k-1}) &=& \int_{\mathds{R}^N} p(\mb{z}_k | \mb{x}_k)p(\mb{x}_k | \mb{z}_{1:k-1}) d\mb{x}_k  \\
\mbox{time update}: p(\mb{x}_{k+1} | \mb{z}_{1:k}) &=& \int_{\mathds{R}^N} p(\mb{x}_{k+1} | \mb{x}_k ) p(\mb{x}_k | \mb{z}_{1:k}) d\mb{x}_k
\eeqna

The {\em posterior distribution} is the primary output from a nonlinear filter, from which standard measures as the minimum mean square estimate $\hat{\mb{x}}_{k|k}^{mms}$ and its covariance $\mb{C}_{k|k}^{(x),mms}$ can be extracted:

\beqna 
\hat{\mb{x}}_{k|k}^{mms} &=& \int \mb{x}_k p(\mb{x}_k | \mb{z}_{1:k}) d\mb{x}_k \\
\mb{C}_{k | k}^{(x),mms} &=& \int \left( \mb{x}_k - \hat{\mb{x}}_k^{mms} \right) \left( \mb{x}_k - \hat{\mb{x}}_k^{mms} \right)^T p(\mb{x}_k | \mb{z}_{1:k}) d\mb{x}_k
\eeqna 

For a linear Gaussian model, the Kalman filter recursion relations provide a solution to this Bayesian filtering problem. But for non-linear or non-Gaussian models no finite-dimensional representations of the posterior density exist, and numerical approximations are needed.

\section{Particle Filter}

If linear Gaussian evolution - observation models are inadequate, the use of the Kalman filter does not result in optimal solutions. In such cases  the posterior density of underlying states $\mb{x}_k$, given the sequence of observations $\mb{z}_{0:k}$ is not analytic, implying that the distribution $p(\mb{x}_k |\mb{z}_{0:k})$ cannot be expressed in a simple form. The application of {\em Monte Carlo} techniques then appears as the most general and robust approach to analyze non-linear system dynamics and/or non-Gaussian state distributions. This is the case despite the availability of the so-called {\em extended} or {\em unscented Kalman filter}, which generally involves a linearization of the problem.

The {\em particle filter} (PF) is also known under various names as there are:

\bit
\item
the bootstrap filter, 
\item
the condensation algorithm,
\item
interacting particle approximations,
\item
survival of the fittest.
\eit

The particle filter approximates the posterior distribution of unobservable states, $p(\mb{x}_k | \mb{z}_{0:k})$, with a discrete density which will be evaluated at a dynamic stochastic grid in this work.

\subsection{Sampling on a dynamic stochastic grid}

The \ac{pf} is based on a direct application of the Bayesian recursion relations and approximates the posterior distribution, $p(\mb{x}_k | \mb{z}_{0:k})$, with a discrete density sampled on a dynamic stochastic grid. This kind of sampling has some noteworthy properties:

\bit 
\item 
A dynamic stochastic grid $\mb{x}^i_k$ changes over time and represents a very efficient representation of the state space.
\item 
The PF generates and evaluates a set $\{\mb{x}^i_{1:k} \}_{i=1}^N$ of $N$ different trajectories rather than current states $\mb{x}^i_k$ only. This affects the time update of the Bayesian recursion relations as follows:

\beqna 
p\left( \mb{x}^i_{1:k+1} | \mb{z}_{1:k} \right) &=& p\left( \mb{x}^i_{1:k} | \mb{z}_{1:k} \right) p\left( \mb{x}^i_{k+1} | \mb{x}^i_{1:k}, \mb{z}_{1:k} \right)  \nonumber \\
&=& w^i_{k|k} p\left( \mb{x}^i_{k+1} | \mb{x}^i_k \right)
\eeqna 

\item 
The new grid is obtained by sampling from 

\beq
p\left( \mb{x}_{k+1} | \mb{z}_{1:k} \right) = \sum_{i=1}^N p\left( \mb{x}^i_{1:k+1} | \mb{z}_{1:k} \right)  = \sum_{i=1}^N w^i_{k|k} p\left( \mb{x}^i_{k+1} | \mb{x}^i_k \right)
\eeq

\eit

Employing {\em importance sampling} amounts to introducing an auxiliary {\em proposal density} $q\left(\mb{x}_{k+1} | \mb{x}_k, \mb{z}_{k+1} \right)$, from which it is easy to sample. The advantage of using an auxiliary proposal distribution can be seen, if one re-writes the Bayesian recursion as follows:

\beqna 
p\left(\mb{x}_{k+1} | \mb{z}_{1:k} \right) &=& \int_{\mathcal{R}^N} p\left( \mb{x}_{k+1} | \mb{x}_k \right) p\left(\mb{x}_k | \mb{z}_{1:k} \right) d\mb{x}_k  \\
&=& \int_{\mathcal{R}^N} q\left(\mb{x}_{k+1} | \mb{x}_k, \mb{z}_{k+1} \right) \frac{p\left( \mb{x}_{k+1} | \mb{x}_k \right)}{q\left(\mb{x}_{k+1} | \mb{x}_k, \mb{z}_{k+1} \right)} p\left(\mb{x}_k | \mb{z}_{1:k} \right) d\mb{x}_k \nn
\eeqna 

Now for each particle a random sample from this auxiliary proposal distribution $\mb{x}^i_{k+1} \sim q\left(\mb{x}_{k+1} | \mb{x}^i_k, \mb{z}_{k+1} \right)$ is drawn, and the {\em posterior probability} is adjusted for each particle with the {\em importance weight} 

\beqna
p\left( \mb{x}_{1:k+1} | \mb{z}_{1:k} \right) &=& \sum_{i=1}^N \frac{p\left(\mb{x}^i_{k+1} | \mb{x}^i_k \right)}{q\left( \mb{x}^i_{k+1} | \mb{x}^i_k , \mb{z}_{k+1} \right)} w^i_{k|k} \delta\left( \mb{x}_{1:k+1} - \mb{x}^i_{1:k+1} \right) \nonumber \\
&=& w^i_{k+1|k} \delta\left( \mb{x}_{1:k+1} - \mb{x}^i_{1:k+1} \right)
\eeqna

Note that the proposal distribution $q\left(\mb{x}^i_{k+1} | \mb{x}^i_k, \mb{z}_{k+1} \right)$ depends on the last state in the particle trajectory $\mb{x}^i_k$ as well as the next measurement $\mb{z}_{k+1}$. The simplest choice of proposal distribution is to use the dynamic model itself, i.~e. $q\left( \mb{x}^i_{k+1} | \mb{x}^i_k, \mb{z}_{k+1} \right) = p\left( \mb{x}^i_{k+1} | \mb{x}^i_k \right)$, leading to $w^i_{k+1|k} = w^i_{k|k}$. 

Thus, the {\em Particle Filter} is a {\em sequential Monte Carlo} technique for the solution of the {\em state estimation problem}. It is a sampling method which approximates the {\em posterior distribution} by making use of its temporal structure.  Again, the key idea is to represent the required {\em posterior density} function by a {\em set of random samples} (particles) with associated weights yielding a particle representation of the posterior density according to

\beq  
p(\mb{x}_k | \mb{z}_{0:k}) \approx \sum_{i=1}^N w^i_{k-1} \delta \left( \mb{x}_k - \mb{x}^i_{k-1}\right)
\eeq 

\noindent where $w^i_k$ represents the weight of particle $\mb{x}^i_k$. Estimates are then computed based on these samples and weights. As the number of samples becomes very large, this Monte Carlo characterization becomes an equivalent representation of the posterior probability function, and the solution approaches the {\em optimal Bayesian estimate}.

Note that for the posterior density a {\em recursive representation} exists according to

\beq  
p\left( \mb{x}_k | \mb{z}_{0:k} \right) = \alpha \cdot p\left( \mb{z}_k | \mb{x}_k \right) \int d\mb{x}_{k-1} p\left( \mb{x}_{k-1} | \mb{z}_{0:k-1}  \right) p\left( \mb{x}_k | \mb{x}_{k-1} \right)
\eeq 

\noindent With a particle representation, this recursive relation simplifies to

\beq  
p\left( \mb{x}_k | \mb{z}_{0:k} \right) \approx \alpha \cdot p\left( \mb{z}_k | \mb{x}_k \right) \sum_{i=1}^N w^i_{k-1} p\left( \mb{x}_k | \mb{x}^i_{k-1} \right)
\eeq 

\noindent How do we create a proper set of particles for representing the posterior distribution $p\left( \mb{x}_k | \mb{z}_{0:k} \right)$? The answer is {\em importance sampling} which will be explained next.

\subsection{Sequential importance sampling}

The technique of {\em importance sampling} is a method for generating {\em fair} samples of a distribution $p(\mb{x}(t))$. Suppose $p(\mb{x}(t))$  is a density from which it is difficult to draw samples, but it is easy to evaluate $p(\mb{x}^i(t))$  for some particular instances $\mb{x}^i$, i.~e. on a grid. Then, an approximation to $p(\mb{x})$ can be given by:

\beqna
p(\mb{x}) &\approx& \sum_{i=1}^N w^i \delta \left( \mb{x} - \mb{x}^i \right) \\
w^i &=& \frac{p(\mb{x})}{q\left(\mb{x}^i \right)}
\eeqna

\noindent where $q(\mb{x}^i)$ denotes any auxiliary {\em proposal distribution}, also called {\em importance density}. In particular, a uniform sampling of the state space could be used but would lead to {\em sample depletion}, or {\em sample degeneracy}, or {\em sample impoverishment}. A more direct auxiliary proposal distribution would be an approximation to the posterior density $p\left( \mb{x}_k | \mb{z}_{1:k}\right)$. This is achieved by {\em re-sampling}, which introduces the required {\em information feedback} from the observations, so trajectories that perform well, will survive the re-sampling.

\subsection{A generic Particle Filter algorithm}

\bit 
\item 
Choose 
\bit 
\item
a {\em proposal distribution} $q\left( \mb{x}_{k+1} | \mb{x}_{1:k}, \mb{z}_{k+1} \right)$, 
\item 
a {\em re-sampling strategy} and 
\item 
the number of particles $N$.
\eit

\item
{\em Initialization}: Generate $\mb{x}^i_1 \sim p_{x_0}, \; i = 1, \ldots ,N$ and set $w^i_{1|0} = \frac{1}{N}$.

\item
{\em Iteration}: for $k = 1, 2, ......$ do
\bit 
\item
{\em Measurement update}: For $i = 1,2,\ldots,N$ compute
\beq 
w^i_{k|k} = \frac{w^i_{k|k-1} p\left(\mb{z}_k | \mb{x}^i_k \right)}{\sum_{i=1}^N w^i_{k|k-1} p\left(\mb{z}_k | \mb{x}^i_k \right)}
\eeq 
\item 
{\em Estimation}: The filtering posterior density is approximated by 
\beq 
\hat{p}\left(\mb{x}_{1:k} | \mb{z}_{1:k} \right) = \sum_{i=1}^N w^i_{k|k} \delta \left(\mb{x}_{1:k} - \mb{x}^i_{1:k} \right) 
\eeq 
and the mean is approximated by 
\beq 
\langle \mb{x} \rangle \approx \sum_{i=1}^N w^i_{k|k} \mb{x}^i_{1:k}
\eeq 
\item 
{\em Re-sampling}: Optionally at each time, take $N$ samples with replacement from the set $\{ \mb{x}^i_{1:k}\}_{i=1}^N$ where the probability to take sample $i$ is $w^i_{k|k} = 1/N$
\item 
{\em Time update}: Generate predictions according to the chosen proposal density
\beq 
\mb{x}^i_{k+1} \sim q\left(\mb{x}_{k+1} | \mb{x}^i_k, \mb{z}_{k+1} \right)
\eeq 
and compensate for the importance weight
\beq 
w^i_{k+1|k} = w^i_{k|k} \frac{p\left(\mb{x}^i_{k+1} | \mb{x}^i_k \right)}{q\left(\mb{x}^i_{k+1} | \mb{x}^i_k, \mb{z}_{k+1} \right)}
\eeq 
\eit
\eit

Note that the algorithm exhibits the fundamental structure of the Bayesian recursion relations as detailed above. Most common forms of PF algorithms combine the weight updates into one equation according to

\beq 
w^i_{k|k} \propto w^i_{k-1|k-1} \frac{p\left(\mb{z}_k | \mb{x}^i_k \right) p\left( \mb{x}^i_k | \mb{x}^i_{k-1}\right)}{q\left(\mb{x}_k | \mb{x}^i_{k-1}, \mb{z}_k \right)}
\eeq

The PF algorithm outputs an approximation of the {\em trajectory posterior density} $p\left(\mb{x}_{1:k} | \mb{z}_{1:k} \right)$. For a filtering problem, the simplest solution would be just to extract the last state $\mb{x}^i_k$ from the trajectory $\mb{x}^i_k$ and use the {\em particle approximation}

\beq 
\hat{p}\left(\mb{x}_k | \mb{z}_{1:k} \right) = \sum_{i=1}^N w^i_{k|k} \delta\left( \mb{x}_k - \mb{x}^i_k \right) 
\eeq 

\noindent However, this is incorrect as in general all paths $\mb{x}^j_{1:k-1}$ can lead to the state $\mb{x}^i_k$. The correct solution, taking all paths leading to $\mb{x}^i_k$ into account, leads to an {\em importance weight}

\beq 
w^i_{k+1|k} = \sum_{j=1}^N w^j_{k|k} \frac{p\left(\mb{x}^i_{k+1} | \mb{x}^j_k \right)}{q\left(\mb{x}^i_{k+1} | \mb{x}^i_k, \mb{z}_{k+1} \right)}
\eeq 

\noindent which replaces the time update of the normalized weights given in the generic \ac{pf} algorithm above. This solution is called the {\em marginal Particle Filter} (mPF) which has many applications in system identification and robotics. Unfortunately, the complexity is now of order $O(N^2)$.

Prediction to get $p\left(\mb{x}_{1:k+m} | \mb{z}_{1:k} \right)$ can be implemented by repeating the time update in the generic PF algorithm $m$ times.

\subsection{Importance re-sampling}

The basic particle filter suffers from sample depletion where all but a few particles  will have negligible weights. Re-sampling solves this problem, but inevitably destroys information and thus increases uncertainty by the random sampling. In this work the  {\em bootstrap PF}, also called \ac{sir}, is employed which applies re-sampling each time. To avoid sample depletion, an auxiliary {\em importance density} needs to be selected appropriately as the prior density $p\left( \mb{x}_k | \mb{x}^i_{k-1} \right)$. 

\subsection{The SIR algorithm}

The \ac{sir} algorithm uses re-sampling at every iteration. It can be summarized in the following steps, as applied to the system evolution from $t_{k-1}$ to $t_k$:

\bit 
\item[Step 1]
\bit 
\item
For $i = 1, \ldots , N$ draw new particles $\mb{x}^i_k$ from the importance density by employing the transition model

\beq  
q(\mb{x}_k ) = \sum_{i=1}^N w^i_{k-1} p\left( \mb{x}_k | \mb{x}^i_{k-1}  \right)
\eeq 

\noindent To do so, choose a random number $r$ uniformly from $[0,1]$ and choose particle $i=r$, then sample from the prior density $p\left( \mb{x}_k | \mb{x}^i_{k-1}\right)$.
\item
Use the corresponding {\em likelihood} to calculate corresponding weights 
\beq 
w^i_k = p\left( \mb{z}_k | \mb{x}^i_k \right)
\eeq 

\noindent The samples $\mb{x}^i_k$, employed above, are fair samples from $p\left( \mb{x}_k | \mb{z}_{0:k-1} \right)$  and re-weighting them accounts for the evidence of the observations $\mb{z}_k$.
\eit
\item[Step 2]
Calculate the total weight $W_k = \sum_{i=1}^N w^i_k $. \\
Normalize the particle weights $w^i_k = W_k^{-1} w^i_k \; \forall i=1, ...., N$
\item[Step 3]
Re-sample the particles by doing
\bit 
\item
Compute the cumulative sum of weights $W^i_k = W^i_{k-1} + w^i_k ; \forall i=1, \ldots ,N, \; W^0 = 0$
\item
Let $i = 1$ and draw a starting point $u_1$ from a uniform distribution $U[0,N^{-1}]$
\item
For $j=1, \ldots ,N$ do the following
\bit
\item
move along the cumulative sum of weights by setting $u_j = u_1 + N^{-1}(j-1)$
\item
while $u_j > W^i$ set $i = i+1$
\item
assign samples $\mb{x}^j_k = \mb{x}^i_k$
\item
assign weights $w^j_k = N^{-1}$
\eit 
\eit
\eit

The resampling procedure just described avoids to have many degenerate particles with vanishing weights but it also leads to a loss of diversity in the sense that the resulting samples may contain many redundant particles. This phenomenon is called {\em sample impoverishment} and is often observed in case of small process noise. In this situation, all particles collapse to a single particle within few instants $t_k$. 

\subsection{Effective number of samples}

An indicator of the {\em degree of depletion} is the effective number of samples, defined in terms of the coefficient of variation $\sigma_{\epsilon_x}$ as

\beq 
N_{eff} = \frac{N}{1+\sigma_{\epsilon_x}^2 (w^i_{k|k})} 
\eeq  

\noindent The effective number of samples is thus at its maximum $N_{eff} = N$ when all weights are equal $w^i_{k|k} = \frac{1}{N}$, and the lowest value it can attain is $N_{eff} = 1$, which occurs when $w^i_{k|k} = 1$ with probability $1/N$ and $w^i_{k|k} = 0$ with probability $(N-1)/N$.

In practical applications this number could be approximated by 

\beq 
\hat{N}_{eff} = \frac{1}{\sum_{i=1}^N \left(w^i_{k|k} \right)^2}
\eeq 

\noindent when we have $1 \le \hat{N}_{eff} \le N$. Again the upper bound $N_{eff} = N$ is attained when all particles have the same weight, and the lower bound $N_{eff} = 1$ when all the probability mass is devoted to a single particle. The resampling condition in the PF can now be defined as $N_{eff} < N_{th}$, and the threshold can for instance be chosen as $N_{th} = \frac{2N}{3}$.

\subsection{Choice of proposal/importance distribution}

The choice of proposal distribution clearly influences the depletion problem.  The most general proposal distribution has the form $q\left(\mb{x}_{1:k} | \mb{z}_{1:k} \right)$. This means that the whole trajectory needs be sampled at each iteration, which in real-time applications is not realistic. But the general proposal can be factorized as

\beq 
q\left(\mb{x}_{1:k} | \mb{z}_{1:k} \right) = q\left(\mb{x}_{k} | \mb{x}_{1:k-1}, \mb{z}_{1:k} \right)q\left( \mb{x}_{1:k-1} | \mb{z}_{1:k} \right)
\eeq 

The most common approximation in applications is to reduce the path $\mb{x}_{1:k-1}$ and only sample the new state $\mb{x}_k$, so the proposal distribution $q\left(\mb{x}_{1:k} | \mb{z}_{1:k} \right)$ is replaced by $q\left(\mb{x}_{k} | \mb{x}_{1:k-1}, \mb{z}_{1:k} \right)$, which, due to the Markov assumption, can be written as

\beq 
q\left(\mb{x}_{k} | \mb{x}_{1:k-1}, \mb{z}_{1:k} \right) \approx q\left(\mb{x}_{k} | \mb{x}_{k-1}, \mb{z}_{k} \right).
\eeq 

The approximate proposal density predicts good values of the current state $\mb{x}_k$ only, not of the whole trajectory $\mb{x}_{1:k}$. For further insight, one needs to discuss the dependence of this proposal distribution  on the \ac{snr}. Here, the \ac{snr} is defined as the ratio of the maximal value of the likelihood and prior, respectively,

\beq 
SNR \propto \frac{max_{x_k} p(\mb{z}_k | \mb{x}_k)}{p(\mb{x}_k | \mb{x}_{k-1})}
\eeq 

\noindent Note that for a {\em linear Gaussian model} this yields 

\beq 
SNR \propto \sqrt{\frac{\det(\mb{C}^{\epsilon_x})}{\det(\mb{C}^{\epsilon_m})}}
\eeq 

\noindent Thus, the \ac{snr} is high if the measurement noise is small compared to the signal noise. Given this measure, one could consider different sampling strategies, but here only the one used in this work is explained.

{\em Likelihood sampling}: 
For {\em medium or high} \ac{snr}, samples are drawn from the {\em likelihood} $p\left( \mb{z}_k | \mb{x}_k \right)$. 

The proposal distribution $q\left(\mb{x}_{k} | \mb{x}^i_{k-1}, \mb{z}_{k} \right)$ can be factorized as follows

\beqna
q\left(\mb{x}_{k} | \mb{x}^i_{k-1}, \mb{z}_{k} \right) \approx p\left(\mb{x}_{k} | \mb{x}^i_{k-1}, \mb{z}_{k} \right) 
&=& p\left(\mb{x}_k | \mb{x}^i_{k-1} \right)  \frac{p\left(\mb{z}_{k} | \mb{x}^i_{k-1}, \mb{x}_{k} \right)}{p\left(\mb{z}_k | \mb{x}^i_{k-1} \right)} \nonumber \\
&=& p\left(\mb{x}_k | \mb{x}^i_{k-1} \right)  \frac{p\left(\mb{z}_{k} | \mb{x}_{k} \right)}{p\left(\mb{z}_k | \mb{x}^i_{k-1} \right)}
\eeqna

\noindent Now consider the case that the likelihood $p\left( \mb{z}_k | \mb{x}_k \right)$ is much more peaky than the prior $p(\mb{x}_k | \mb{x}_{k-1})$ and if it is integrable in state space, $\mb{x}_k$, then one can set

\beq  
d^2(\mb{x}_n, \mb{x}_m)
p\left(\mb{x}_k | \mb{x}^i_{k-1}, \mb{z}_k \right) \propto p\left(\mb{z}_k | \mb{x}_k \right)
\eeq

\noindent Thus, a suitable proposal distribution for the {\em high} \ac{snr} case is based on a {\em scaled likelihood function}

\beq 
q\left(\mb{x}_k | \mb{x}^i_{k-1}, \mb{z}_k \right) \propto p\left(\mb{z}_k | \mb{x}_k \right)
\eeq 

\noindent This choice then yields the following weight update

\beq 
w^i_{k|k} = w^i_{k-1|k-1} p\left(\mb{x}^i_k | \mb{x}^i_{k-1} \right)
\eeq 

\noindent Sampling from the {\em likelihood} requires that the likelihood function $p\left(\mb{z}_k | \mb{x}_k \right)$ is integrable with respect to $\mb{x}_k$. This is not the case when $N > L$, i.~e. the number of unobservable states is larger than the number of observations (measurements). The interpretation in this case is that for each value of $\mb{z}_k$, there is an infinite-dimensional manifold of possible $\mb{x}_k$ to sample from, each one equally likely.

\section{Resampling Schemes}
The method of sampling from the {\em likelihood} has the drawback that it becomes unstable as $k$ increases. There is a discrepancy between the weights, therefore the algorithm can be stabilized by performing a {\em re-sampling} sufficiently often. The weighted approximate density is modified by each re-sampling step. This is done by eliminating particles with low weights and multiplying particles with an important weight. The filtering posterior density then becomes:
\beq 
\hat{p}\left(\mb{x}_{0:k}, \mb{z}_{1:k} \right) = \sum_{i=1}^N \frac{n_i}{N} \delta(\mb{x}_{0:k} - \mb{x}_{0:k}^i)
\eeq 
where $n_i$ is the number of copies of particle $\mb{x}_{0:k}^i$. In the following, the three main {\em re-sampling schemes} will be described.

\subsection{Multinomial Resampling}
The {\em multinomial resampling} scheme generates $N$ ordered uniform random numbers on the interval $(0,1]$. The new particle set is selected according to the multinomial distribution \cite{Douc2005}.

\begin{align*}
u_k &= u_{k+1}\tilde{u}_k^{\frac{1}{k}}, \ with\ \tilde{u}_k \sim U[0,1)  \\
u_N &=\tilde{u}_N^{\frac{1}{N}}\\
\mb{x}_k &= x( F^{-1}(u_k))\\
&= x_i \ with\ i\ so\ that\ u_k \in \big[ \sum_{s=1}^{i-1} \omega_s, \sum_{s=1}^{i} \omega_s \big),
\end{align*}

\noindent where the generalized inverse of the cumulative probability distribution is $F^{-1}$. The weights are normalized.

\subsection{Systematic resampling}
With {\em systematic re-sampling}, the particle set is selected according to the multinomial distribution of $N$ generated ordered numbers $u_k$.

\beq 
u_k= \frac{(k-1)+\tilde{u}}{N},\ with\ \tilde{u}\sim U[0,1)
\eeq 

\subsection{Residual resampling}
With {\em residual re-sampling}, the variance can be decreased. We define:

\begin{align*}
N^i= \lfloor n \omega^i \rfloor + \bar{N}^i,\ \mbox{for} \ i=1,\ldots ,m,
\end{align*}

\noindent where the $\bar{N}^1, ..., \bar{N}^n$ are multinomially distributed, and $\lfloor \rfloor$ means the integer part. The multinomial distribution $M(n-R; \bar{\omega}^1,...,\bar{\omega}^n)$ has the following arguments:

\begin{align*}
R &= \sum_{i=1}^m \lfloor n\omega^i \rfloor \\
\bar{\omega}^i &= \frac{n\omega^i - \lfloor n \omega^i \rfloor}{n-R}, \ i=1,...,m
\end{align*}

\noindent The multinomial counts $\bar{N}^1,..., \bar{N}^n$ are generated in the same way as in the {\em multinomial re-sampling} described above.

\chapter{Time-Integrated Activity Map (TIA)}

In this chapter the four \ac{spect}-images as shown in figure \ref{fig:patient_dat} will be integrated voxelwise to obtain the number of decays per voxel. The activity values will be denoised by the particle filter. After the denoising, a mono-exponential function is fit to the data and integrated over time.

\section{Particle Filter Method}

For dosimetry it is more important to have a patient specific model than an on-line tracking method. Therefore all \ac{spect} images were used to estimate a state evolution model. First of all, \ac{voi}s were segmented in a \ac{ct} image by a medical expert. To this contour a safety margin of 2 voxels was added to account for partial volume effects as well as deviations from the registration or inter-organ movements. By summing up the activity values of all voxels enclosed by the contour, an {\em organ-specific total activity} was obtained for each time point. An additional time point was added at $t_k = 600\ h$, at which the specific activity was set to $0\ Bq/ml$ deliberately. The time-dependent activity (see figure \ref{fig:estimated_time_activity_curve}) was expressed by a parametrized analytic function which was fit to the observed time-dependent activities according to

\beq
\ln \left( \frac{a(t)}{a(0)} \right) = - \left(\frac{ln(2)}{T_{1/2}}\cdot t \right) = - \frac{t}{\tau_{eff}} 
\eeq

\noindent
Note that the effective lifetime 
\beq 
\tau_{eff} = \left( \frac{1}{\tau_{phys}} + \frac{1}{\tau_{biol}} \right)^{-1} = \frac{T_{1/2}}{\ln 2}
\eeq 

\begin{figure}[!htb]
	\centering
	\includegraphics[width=0.5\textwidth]{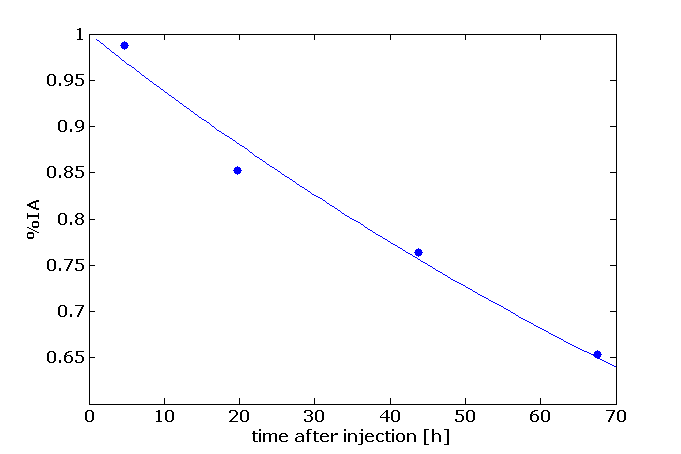}
	\caption{Time activity curve for one kidney: $f(0) = 1 \%IA, T_{1/2} = 77.9 h$.}
	\label{fig:estimated_time_activity_curve}
\end{figure}

is composed of the physical lifetime of the radionuclide and its biological lifetime due to the metabolic turnover of the radio-pharmaceutical. While the biological half-life of the isotope should be identical across an organ, the amplitude $a(0)$ of the \ac{tac} is different for every voxel. Note that $a(0)$ is different from the initially injected activity $A_0$ and generally unkown.

\subsection{State evolution model (SEM)} 
For each voxel of the kidney an own \ac{sem} is computed. At the beginning, the four activity values of the same voxel plus an additional value of $a(t_k) = 0 Bq/ml$ at $t_k = 600\ h$ are approximated by a mono-exponential decay.  Hereby, the half-life is assumed to be equal to the previously computed half-life of the whole kidney. From this fit, the amplitude $\langle a(0) \rangle$ of the state evolution model can be obtained. Moreover, the inevitable noise on the data is modeled as Gaussian white noise with an amplitude corresponding to $0.1 \langle a(0) \rangle$. 

\subsection{Observation model (OM)} 
The \ac{om} $\mb{h}$ represents the distribution of acquisition times of the \ac{spect} image and is modeled by a Normal distribution around the estimated particle.

\subsection{Similarity measure}
The aim of the voxelwise integration of the activity is to obtain the number of nuclear disintegrations $N^{Lu}_n$ for each voxel $n \in S$ according to:

\beq 
N^{Lu}_n = \int_0^{t_K} a_n(t') dt' \approx \sum_{k=1}^K a_n(t_k) \cdot \Delta t
\eeq 

\noindent where $K = 5$ in this study. The activity distribution is assumed to be identical for each voxel, hence it can be estimated for each organ at every  time point.  This fact can be explored to find a good parameter configuration.  As similarity measure the \ac{kl} was used. An averaged histogram of the $k = 4$ time points was calculated, for which the histograms were centered. A histogram of the voxel activities $h(\langle a_n \rangle)$ was calculated, where $\langle a \rangle_n = (1/K)\sum_k a_n(t_k)$ denotes the activity of voxel $n$ averaged over the four time points. The similarity between different histograms has been computed employing the \ac{kl}.

\begin{figure}[!htb]
	\centering
	\includegraphics[width=0.49\textwidth]{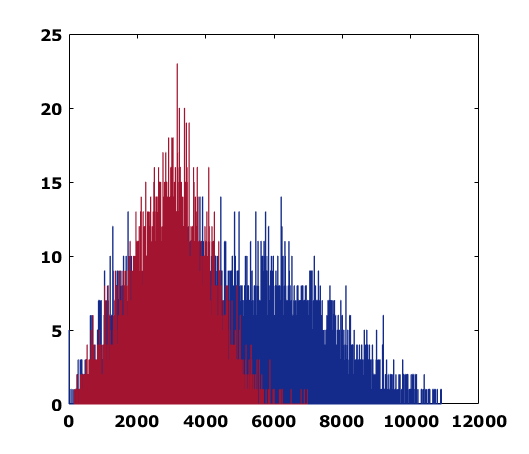}
	\includegraphics[width=0.49\textwidth]{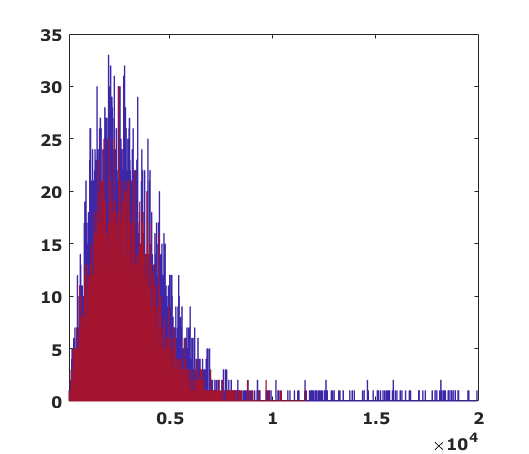}
	\caption{Shown are two histogramms of voxelactivities: Left side with a Kullback-Leibler divergence of $0.67$ and on the right side with $0.12$.}
	\label{fig:estimated_time_activity_curve}
\end{figure}
\section{Simple Fit Method}

Instead of fitting the activity values as smoothed by the particle filter ('particle fit'), one can use another approach to calculate a voxel-wise decay map. One can fit a mono-exponential function to the four activity values per voxel. In the follwoing this method is called 'simple fit'.

For each method the \ac{tia} in a specific source region $S$ was calculated by integrating, for every voxel, the fit function over time and sum over all voxels.

\beqa
TIA_{S} &=& A_0 \sum_{n \in S} \int_0^{\infty} \cdot e^{- \frac{t}{\tau_{eff,n}}} dt  \nonumber 
\eeqa 

where $S$ is the source region, $n$ indicates one voxel, $A_0$ denotes the injected activity and $\tau_{eff,n}$ the half-life, which can differ between the voxels for both, the 'simple fit' and 'particle fit' method. Note that the approximation $a_n(t_0) \approx A_0 \forall \; n \in S$ accounts for the fact that the complex metabolic processes which distribute the radioactive pharmaceutical in the body is generally unknown. Finally, also the distribution of half-lives across all voxels in the \ac{voi}s was calculated.

\chapter{Results}

\section{Parameter evaluation}

Finding a good parameter configuration, which yields similar histograms of the number of decays within an organ, and which also yields a good match to the measured activity distribution of all activity values within a \ac{voi}, the standard deviations for the state evolution model and the observation model were considered adjustable parameters. 
The configurations were computed for all $26$ patients, and for each configuration, the histogram of the number of decays per voxel was determined and compared with the averaged activity histogram. The \ac{kl} was computed for all configurations. The $\sigma_{obs}$ was varied between $0.35 \le \sigma_{obs} \le 0.99$. The considered standard deviations are rather large because the estimated precision of the activity measurement through \ac{spect} imaging is in this range according to \cite{ritt2011technik}.

In the following table, the \ac{kl}s, averaged over all patients, are collected:

\begin{table}[!htb]
\caption{Kullback - Leibler divergences, averaged over the entire patient cohort}
\begin{center}
\begin{tabular}{|c|c|c|c|c|} \hline
model$ \diagdown$ observation	&	0.35	    & 0.5		& 0.75		& 0.99 \\ \hline
0.01				&  0.12        & 0.17     & 0.17     & 0.17 \\
0.10 				&  0.16        & 0.17     & 0.17     & 0.17 \\
0.25				&{\cb 0.16}    & 0.16     & 0.16     & 0.16 \\
0.35				&  0.58        & 0.54     & 0.48     & 0.46 \\
\hline
\end{tabular}
\end{center}
\end{table}

The values in each row are very similar to each other, indicating that the \ac{kl} is rather robust against uncertanties in the observation model. The smallest value for the \ac{kl} is marked in blue.

\begin{table}[!htb]
\caption{Relative deviations of the sum of voxelwise integrated activities from the corresponding whole organ integrated activity in percent}
\begin{center}
\begin{tabular}{|c|c|c|c|c|}\hline
model $\diagdown$ observation	&	0.35	& 0.5		& 0.75		& 0.99 \\ \hline
0.01				& $-34.2 \%$& $-32.6\%$	& $-32.6\%$ & $\textcolor{blue}{-32.6\%}$ \\
0.10 				& $-35.8 \%$& $-33.7\%$ & $-33.4\%$ & $-33.4\%$ \\
0.25				& $-44.4 \%$& $-40.3\%$ & $-39.0\%$ & $-38.5\%$ \\
0.35				& $-64.4 \%$& $-59.3\%$	& $-56.3\%$	& $-55.9\%$ \\ \hline
\end{tabular}
\end{center}
\end{table}

As a second decision criterion for an optimal parameter configuration, the relative deviation of the total number of decays  estimated either by integrating the time-dependent activity at every voxel  and then summing over all voxels within the \ac{voi} or by integrating the whole organ (here the kidney) activity over time was calculated.

In the following table the resulting relative deviations are collected. The deviations for the total number of decays are large because of a high statistical uncertainty in the voxel activity values. To correct for these large deviations, the voxelwise integrated activity map can be normalized with the total number of decays per organ. For the following results, we choose the parameters as $\sigma_{observation}=0.35$ and $\sigma_{model}=0.10$. 

\begin{figure}[!htb]
	\centering
	\includegraphics[width=0.8\textwidth]{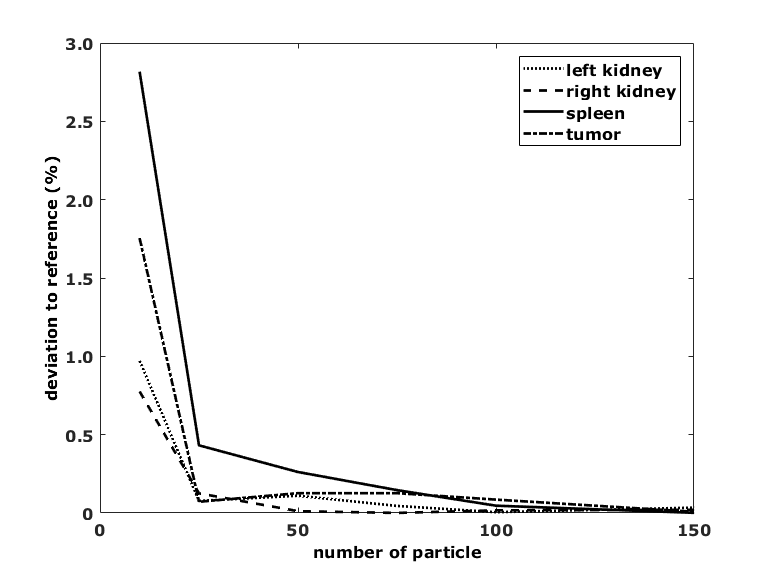}
	\caption{Variation of the number of particles.}
	\label{fig:number_of_particle}
\end{figure}

To obtain stable results for the single activity values, the number of particles for each time step was varied. In figure \ref{fig:number_of_particle} the determined difference between the total number of decays per organ for different numbers of particles and the total number of decays for $2000$ particles is illustrated. With only $150$ particles, the difference to the reference value almost disappears.

Moreover, while applying the particle filter, different resampling schemes were evaluated, but there was no difference in the total number of decays per organ. Therefore we decided to perform no resampling. The possibility to loose too much diversity is very low with only four time steps.

\section{Voxelwise TAC}
\begin{figure}[!htb]
	\centering
	\includegraphics[width=0.49\textwidth]{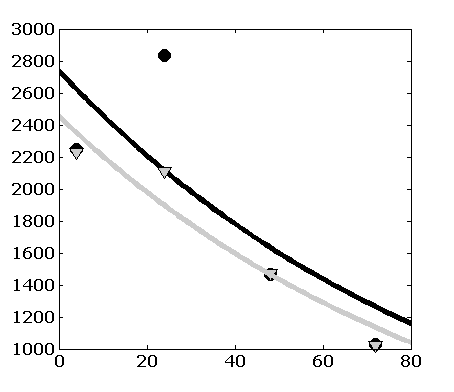}
	\includegraphics[width=0.49\textwidth]{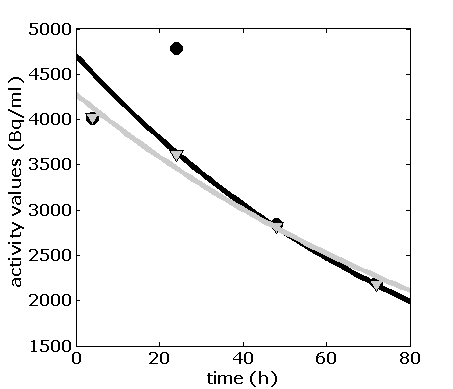}
	\caption{Time activity curves for two different voxels at four time steps.}
	\label{fig:one_voxel}
\end{figure}

For each voxel, a state evolution model was defined through four measured time points. According to the state evolution model and the observation model, particles were generated and an estimation of the true activity value was obtained through its likelihood. An analytic function was then adapted to  either the four estimates of the true activity values or the observed activity values. In figure \ref{fig:one_voxel}, the actual measurements and the estimates of the state evolution model as represented by properly adapted analytic functions, whereby the adaptation to the observed values is drawn in black and the adaptation to the estimates of the true activity values is colored in grey. Through particle filtering, the outlier at the second time point could be eliminated very well.

\section{Whole patient}
\begin{figure}[!htb]
	\centering
	\includegraphics[width=0.8\textwidth]{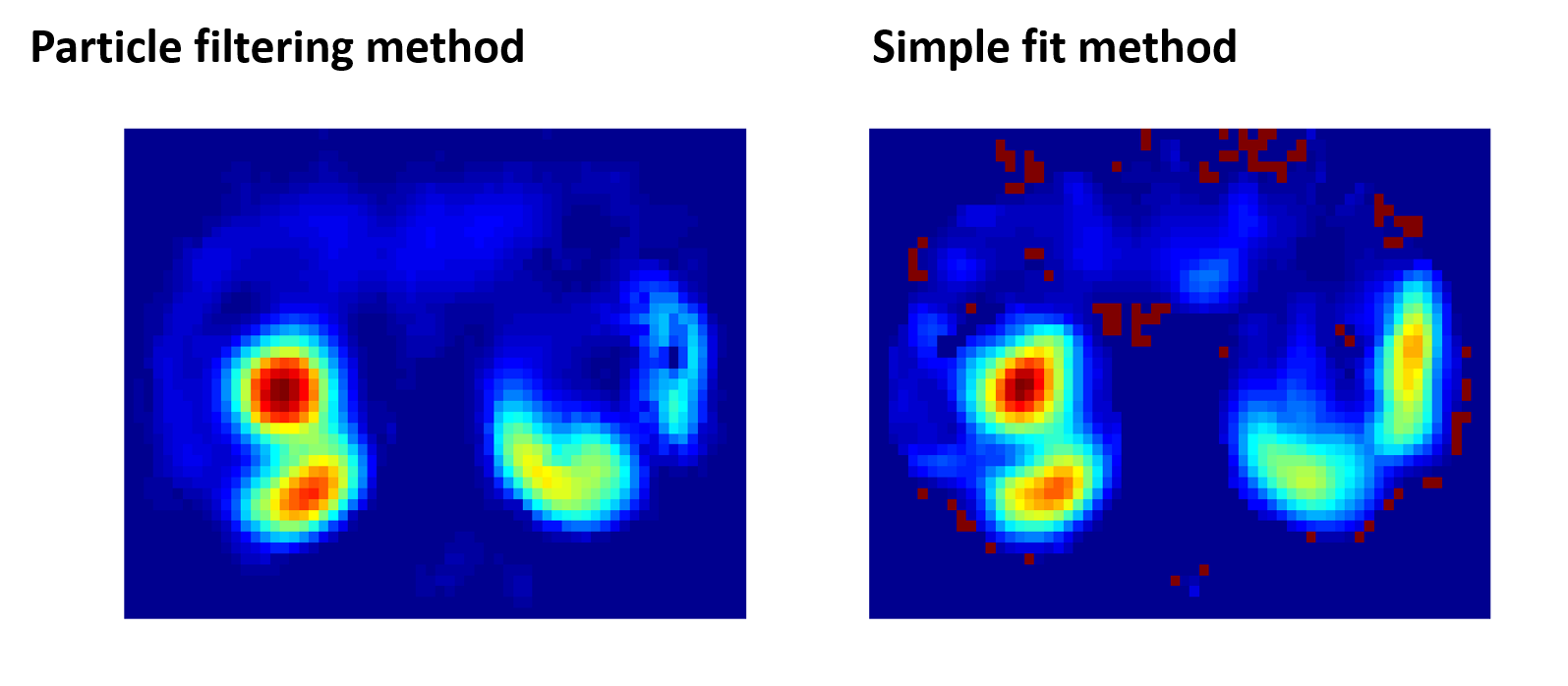}
	\caption{One slice of the results for the particle filtering method (left) and the simple fit method (right).}
	\label{fig:results_particle_simple}
\end{figure}

In figure \ref{fig:patient_dat}, one slice of the \ac{spect} images at four time points for one patient have been illustrated.  Now the voxelwise results for the integration of the \ac{tac}s are shown in figure \ref{fig:results_particle_simple}. The red voxels in the right image, showing the results from the simple fit method, are those with a diverging integral, the calculated number of decays is equal to infinity.

\begin{figure}[!htb]
	\centering
	\includegraphics[width=0.5\textwidth]{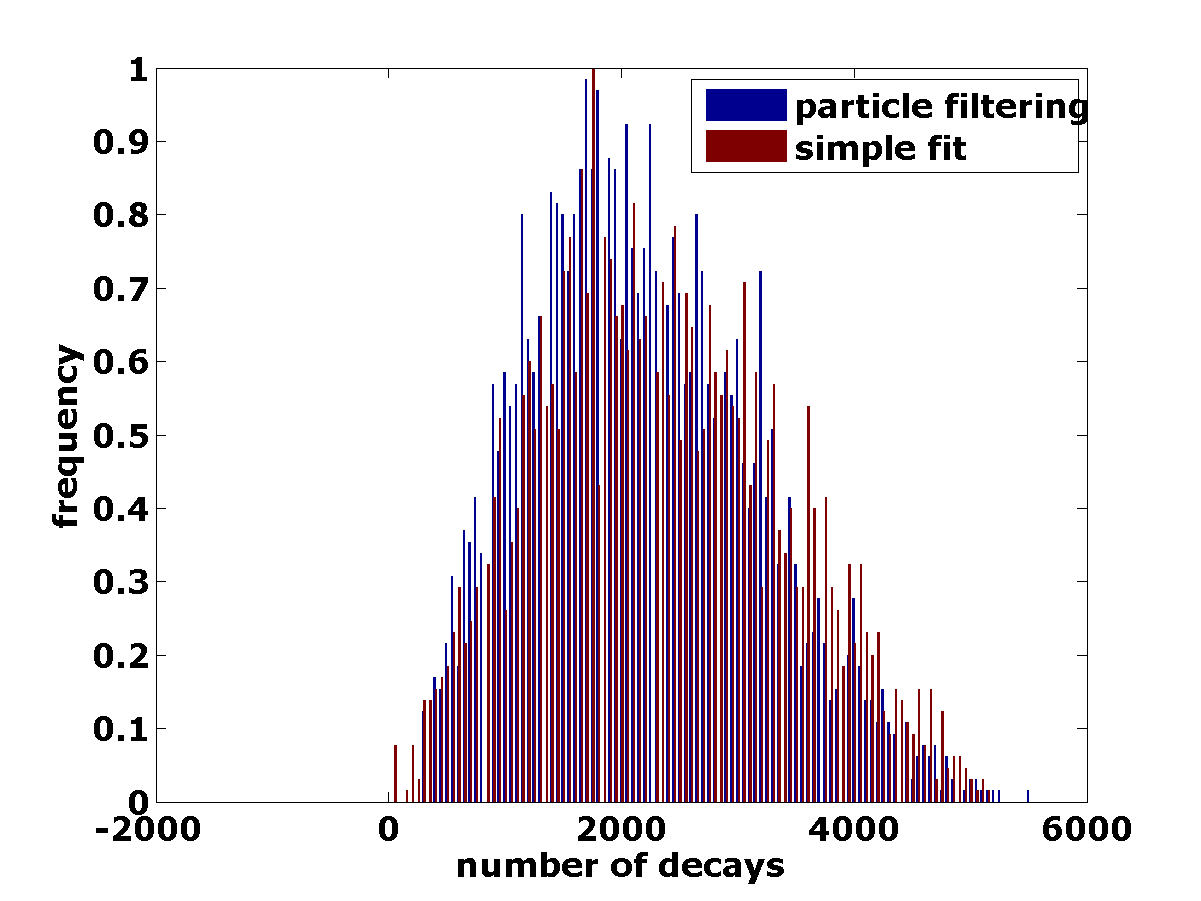}
	\caption{Histograms of number of decays per voxel for the particle filter method (blue) and the voxelwise simple fit of the four activity values (red).}
	\label{fig:hist_decays}
\end{figure}

 The voxelwise integration of the particle filtered activity values was performed for the whole patient. For the segmented organs (kidney, spleen and tumor), a corresponding half-life was calculated and used for the state evolution model. For the voxels outside the \ac{voi}s, an over all voxels from the same tissue class averaged half-life was calculated and used for the model.
For each \ac{voi}, a histogram of the number of decays per voxel was obtained and is shown in figure \ref{fig:hist_decays}. A second histogram is also shown, were the number of decays per voxel is presented, which is calculated by a simple voxel-by-voxel integration. Both histograms are almost indistinguishable.

\begin{figure}[!htb]
	\centering
	\includegraphics[width=0.5\textwidth]{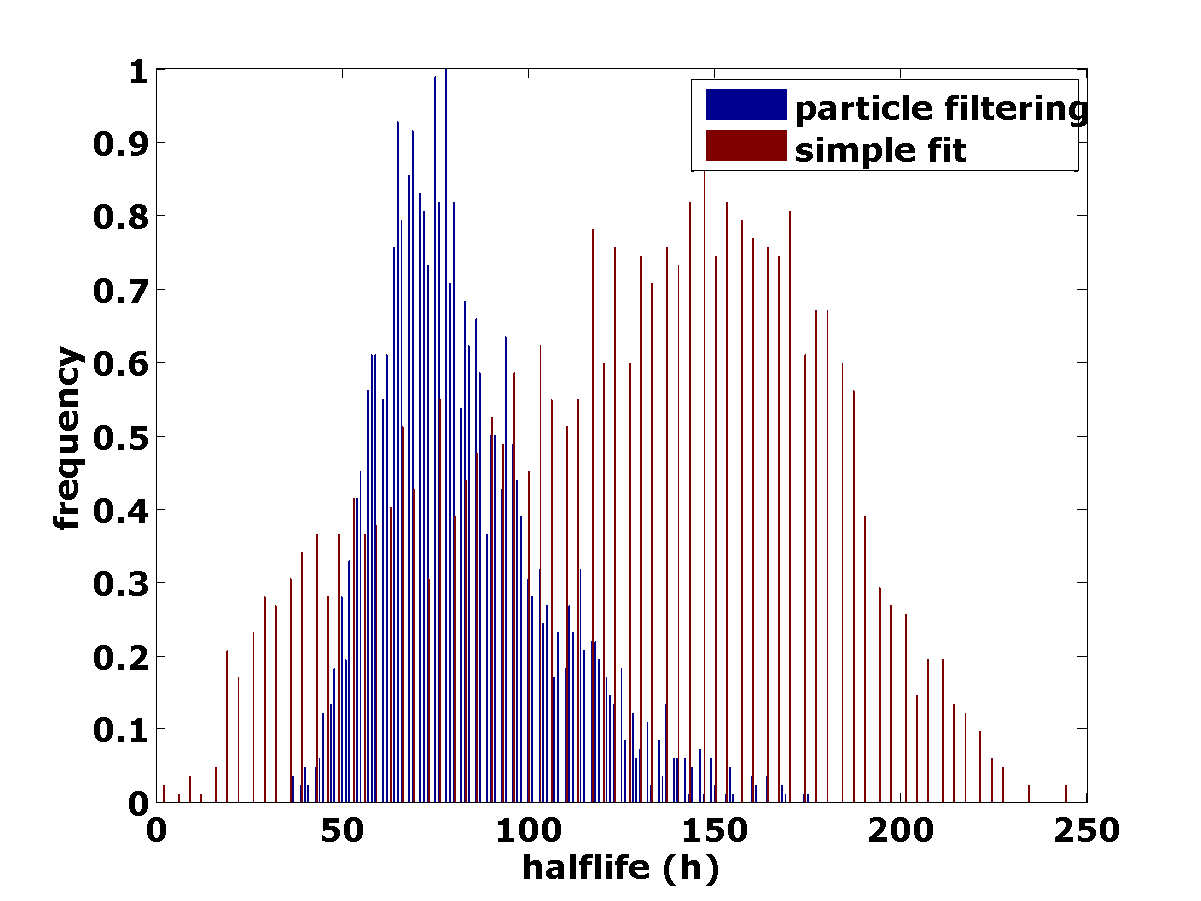}
	\caption{Histograms of halflives per voxel for the particle filter method (blue) and the voxelwise simple fit of the four activity values (red). The halflife for the whole organ is $81.6h$.}
	\label{fig:hist_halflife}
\end{figure}

Considering the histograms of all voxel-specific half-lives, which were used in the analytic function for the determination of the \ac{tia}, the particle filter method yielded histograms centered at the half-life from the state evolution model, while the simple fit method very often yielded  half-lives larger than the physical half-life of Lutetium (see figure \ref{fig:hist_halflife}). Also the widths of the two distributions differ considerably. The much narrower width of the distribution for the particle filter method depends on the $\sigma$ parameters of the \ac{sem} and the \ac{om}.

\chapter{Discussion}

In this work, we proposed a method for processing voxel-wise \ac{tac}s prior to fitting them to mono-exponentials and subsequently  integrating them temporally. The \ac{tac}s were obtained by sequential, quantitative \ac{spect/ct} imaging for dosimetry of $26$ patients under Lu-177 targeted radiotherapy. The proposed pre-processing method is a \textit{Particle Filter}, which belongs to the class of sequential Monte-Carlo methods. The aim of the method is a de-noising of the observed \ac{tac}s. The goals were the following:
\bit 
\item 
The whole-organ \ac{tia} map, as obtained on basis of the filtered data, should be as similar as possible to the conventional evaluation. The latter obtains TIA maps by fitting and integrating the organ-averaged time-activity curve.
\item 
The histograms of the \ac{tia} maps, resulting from the PF, should resemble the \ac{tia}-histograms, as initially estimated from \ac{spect} images.
\eit

We have shown that \textit{Particle Filters} are suitable for a reduction in the variance of the effective half-lives of voxels in kidneys, spleen, and tumors. Most notably, the number of voxels which exhibit half-lives longer than the physical half-life of the radioisotope is significantly reduced.

For two voxel-specific methods to estimate whole organ \ac{tia}s, namely \textit{particle fit} (PFF) and \textit{simple fit} (SF), we quantified deviations to whole-organ \ac{tia}s estimated by time-integration of the whole organ activity. 

When comparing the resulting half-lives, we found that application of particle filtering results in a strong reduction of variance, compared to SF. As intended, the voxel-specific half-lives for PFF were distributed around the half-life used in the state evolution model. Importantly, the number of voxels which exhibit a longer-than-physical half-life was greatly reduced with PFF, compared to SF. In summary, the PFF method does not fix the half-life to a single value, rather it exhibits a  narrow distribution and reduces the number of implausible half-lives to almost zero in our study.

Unfortunately, literature for comparing different methods for fitting voxel-specific \ac{tac}s is scarce so far. Thus, no generally accepted standard-method exists. Guidelines recommend either not modelling at all, rather performing trapezoidal integration of \ac{tac}s, or modelling by exponential functions \cite{vicini2008, siegel1999}, although these recommendations are usually in the context of whole-organ dosimetry.

In the literature of voxel-based dosimetry in targeted radiotherapy, most groups applied no modelling at all for the time period covered by imaging, and mono-exponential extrapolation after the last imaging \cite{Sgouros2004,Prideaux2007,Wilderman2006,grassi2015quantitative,Sarrut20173DA}. Jackson et al. \cite{jackson2013automated} used tri-exponential modelling of the \ac{tac}, but applied regularization in order to limit the number of voxels with implausible \ac{tia}. In the study of Kost et al. \cite{kost2015}, the user could choose between mono- or bi-exponential modelling of pharmacokinetics. However, the focus of their study was not on the comparison of the different models and consequently they did not report on this.

Marcatili et al. \cite{Marcatili2013} implemented an algorithm which automatically selects the best model for each voxel, either pure mono-exponential decay or linear uptake followed by mono-exponential decay, and no modelling and trapezoidal integration as fall-back if modelling fails. They did not provide a systematic comparison between the different models but stated that the fall-back was used in less than $5\%$ of voxels for their application (I-124 \ac{pet}). However, it could be hypothesized that for \ac{spect} imaging of typical Lu-177 activities, the image noise would be comparably higher than for \ac{pet}, which potentially could affect fitting of per-voxel \ac{tac}s negatively.

To the best of our knowledge, the study of Sarrut et al. \cite{Sarrut2017VBM} is the only study that systematically compared different models for voxel-wise \ac{tac} fitting. They focused on an algorithm called VoMM, that automatically selected the best model for fitting the \ac{tac} of individual voxels. The models considered were mono- or bi-exponential functions with variable degrees-of-freedom of the fit ranging from 2 to 4 parameters. They compared the results obtained by VoMM and by using fixed models for all voxels in a numerical experiment with varying noise levels and also in one dosimetry patient. In any case, 6 imaging time points were used. For the numerical experiment, they evaluated the Root Mean Squared Error between ground-truth and modelled \ac{tia}s and also the percentage of voxels which could not be fit successfully ($\%$FF). They found that the Root Mean Squared Error and $\%$FF was lower for VoMM as compared to the fixed models. Notably, they also showed that a low count number and thus high noise level in a voxel's \ac{tac} would result in an increased Root Mean Squared Error and $\%$FF. For their patient data, they found that $\%$FF was lower for the VoMM than for most fixed models. However, one fixed model did have an even lower $\%$FF.

Altogether, literature on this topic shows that voxel-wise fitting of \ac{tac}s, subsequent calculation of \ac{tia}s, and thus voxel-specific dosimetry is feasible. However, it also turns out that especially the \ac{tac} fitting is most likely negatively affected by noise. Most groups apply some technique to either circumvent modelling of especially the early part of the TACs by using numerical integration or implement some kind of regularization in order to limit the bias introduced by voxel noise. Most groups did only assess if voxel-wise fitting could be done successfully from an algorithmic point of view, but did not evaluate if a successful fit resulted in feasible \ac{tia}s. Based on our experience and from our results obtained by the simple fit method, we suspect that many voxels' \ac{tia}s, although obtained by successful fitting, are still incorrect. This problem could be alleviated by application of \textit{Particle Filters}, which could be understood as de-noising based on a plausible assumption/model. Altogether, we think that particle filters offer an interesting alternative to methods reported from literature and should be further evaluated in forthcoming studies.

\section{Conclusion}

In this study, the challenge was to calculate the number of decays per voxel from the \ac{spect}-images. Therefore, a statistical method called particle filter was used. With this method the voxelwise noise of the \ac{spect} images could be removed by matching the measuremenst to a physical model of the radioactive decay. This method yields a homogeneous distribution of the decays per voxel by taking into account different organs. The particle filter method was compared to a simple fit method. The latter generally resulted in inhomogeneous decay maps where certain voxels contained diverging numbers of decays. For $26$ patients suffering from a prostate cancer or a neuroendocrine tumor, the corresponding decay maps could be determined reliably.


\bibliographystyle{plain}
\bibliography{Diss}
\end{document}